\documentclass[usenatbib]{mnras}

\usepackage{mathptmx}
\usepackage{txfonts}
\usepackage{animate}
\usepackage{comment}
\usepackage{gensymb}
\usepackage{multirow}

\usepackage[T1]{fontenc}

\DeclareRobustCommand{\VAN}[3]{#2}
\let\VANthebibliography\thebibliography
\def\thebibliography{\DeclareRobustCommand{\VAN}[3]{##3}\VANthebibliography}

\usepackage{graphicx}	
\usepackage{amssymb}	
\usepackage{booktabs}

\usepackage{array}
\newcolumntype{M}[1]{>{\centering\arraybackslash}m{#1}}
\usepackage[justification=centering]{caption}

\usepackage[normalem]{ulem}

\usepackage[dvipsnames]{xcolor}

\emergencystretch=\maxdimen
\title[Detecting planetary mass companions with \textit{JWST} interferometry]{Detecting planetary mass companions near the water frost-line using\\ \textit{JWST} interferometry}
\author[S. Ray et al.]{
\newauthor
Shrishmoy Ray$^{1}$\thanks{E-mail:S.Ray2@exeter.ac.uk},
Sasha Hinkley$^{1}$,
Steph Sallum$^{2}$,
Mariangela Bonavita$^{3,4}$,
\newauthor
Vito Squicciarini$^{5,6}$,
Aarynn L. Carter$^{7}$,
and Cecilia Lazzoni$^{1}$
\\
\\
$^{1}$Astrophysics Group, University of Exeter, Physics Building,
Stocker Road, Exeter, EX4 4QL, UK\\
$^{2}$Department of Physics and Astronomy, University of California, Irvine, 4129 Frederick Reines Hall, Irvine, CA 92697-4575, USA\\
$^{3}$Institute for Astronomy, University of Edinburgh, Royal Observatory, Blackford Hill, Edinburgh, EH9 3HJ, UK\\
$^{4}$School of Physical Sciences, Faculty of Science, Technology, Engineering and Mathematics, The Open University, Walton Hall, Milton Keynes, MK7 6AA\\
$^{5}$Department of Physics and Astronomy ‘Galileo Galilei’, University of Padova, Via dell’Osservatorio 3, I-35122 Padova, Italy\\
$^{6}$INAF – Osservatorio Astronomico di Padova, Vicolo dell’Osservatorio 5, I-35122 Padova, Italy\\
$^{7}$Department of Astronomy and Astrophysics, University of California, Santa Cruz, Santa Cruz, 1156 High Street, Santa Cruz, CA 95064, USA
}

\date{Accepted 2022 November 17. Received 2022 November 16; in original form 2022 May 16}

\pubyear{2022}

\begin{document}
\label{firstpage}
\pagerange{\pageref{firstpage}--\pageref{lastpage}}
\maketitle

\begin{abstract}
 \textit{JWST} promises to be the {most versatile} infrared observatory for the next {two} decades. The \textit{Near Infrared and Slitless Spectrograph (NIRISS)} instrument, when used in the Aperture Masking Interferometry (AMI) mode, will provide an unparalleled combination of {angular resolution and} sensitivity compared to any existing observatory at mid-infrared wavelengths. Using simulated observations in conjunction with evolutionary models, we present the capability of this mode to image {planetary mass} companions around nearby stars at small orbital separations near the circumstellar water frost-line for members of the young, kinematic moving groups $\beta$ Pictoris, TW Hydrae, as well as the Taurus-Auriga association. We show that for appropriately chosen stars, \textit{JWST/NIRISS} operating in the AMI mode can image sub-Jupiter companions near the water frost-lines with  ${\sim}68\%$ confidence. Among these, M-type stars are the most promising. We also show that this \textit{JWST} mode will improve the minimum inner working angle by as much as ${\sim}50\%$ in most cases when compared to the survey results from the best ground-based {exoplanet direct imaging} facilities {(e.g. \textit{VLT/SPHERE})}. {We also discuss how the NIRISS/AMI mode will be especially powerful for the mid-infrared characterization the numerous exoplanets expected to be revealed by \textit{Gaia}. When combined with dynamical masses from \textit{Gaia}, such measurements will provide a much more robust characterization of the initial entropies of these young planets, thereby placing powerful constraints on their early thermal histories. }



\end{abstract}




\section{Introduction}

\label{sec:Intro}
High contrast imaging of nearby circumstellar environments is the only technique that provides sensitivity to planetary mass companions (PMCs hereon) at wide orbital separations \citep[e.g.][]{ 2016Bowler}, and hence will extensively map out the outer architectures of planetary systems through the coming years with the advent of next generation telescopes. Previous efforts have led to successful detection of {wide-separation} PMCs \citep[e.g.][]{hr87992,BetaPicPlanet,2017Chauvin, bkg20} and numerous scattered-light images of disks  \citep[e.g.][]{2017Matthews, 2017Mili, 2020Esposito, hml21}. Since this technique preferentially observes young stars, it is also exceptionally well-positioned to place valuable constraints on competing models of planet formation and migration that describe the early dynamical and thermal evolution of planets  \citep[][]{2009PlanetFormation,2009Migration,Marleau2014, Wallace2021}.

{Careful analysis of r}ecent direct imaging (DI, hereon) exoplanet surveys \citep[][]{2019Nielsen, Wagner:2019ApJ, Vigan2021A&A} indicate that numerous lower mass PMCs exist at wide orbital separations (tens to hundreds of au). {Specifically, extrapolating} mass distribution power laws derived by the Gemini Planet Imager (\textit{GPI} hereon) survey \citep[][]{2019Nielsen} demonstrates that an abundance of $0.1{-}1.0\,\rm{M\textsubscript{Jup}}$ planets should be hosted by stars with masses $0.2{-}5.0\,\rm{M_\odot}$. Microlensing efforts \citep[][]{2021Poleski} are also consistent with this prediction, providing statistical {evidence} for the existence of ${\sim}1.5$ ice giant planets {(${\lesssim}1\,\rm{M_{Jup}}$) \textit{per star}} at separations $5{-}15\,\rm{au}$. Detecting {an abundance of such} companions would be significantly valuable for {evaluating the early thermal histories of giant planets, and possibly assigning} populations of planets to formation mechanism models based on {the  accretion of solids in a protoplanetary disk \citep[][]{1996Pollack} {or} the formation of a planet triggered by an instability within the disk \citep[e.g.][]{2010Kratter}}.

Even with all these remarkable observational accomplishments and the development of state-of-the-art models mapping planet formation histories \citep[e.g.,][]{2016Mordasini,2017MOrdasini,mmb22}, several fundamental questions still remain unanswered. The exact details of the gas-giant planet formation process, as well as the physical {and thermodynamic} conditions of newly formed planets, remain unclear \citep[][]{Marleau2014}. This is reflected in the fact that models of luminosity evolution of planets {vary} by orders of magnitude at the youngest ages \citep[][]{2008Fortney, 2012Spiegel}. 
Early entropy conditions, being the single best route towards {enlightening the complex early physics of planet formation} \citep{Marleau2014,Wallace2021}, {still remain largely unconstrained}. 
Provided that we can access {their orbital locations, obtaining luminosity measurements of numerous young planets with the goal of measuring their entropy will be a major focus for DI searches going forward.}



{Coronagraphic} DI surveys in the last $10{-}15$ years \citep[e.g.][]{2015Chauvin, 2016Galicher, 2017Vigan, Vigan2021A&A}, have had a low rate of detection of companions around host stars, {returning a number of companions that  is insufficient to statistically place constraints on planetary formation models as well as models of the early entropy of planets}. 
This poor detection rate of DI planets may be due to the fact that the recent studies {\citep[e.g.,][]{2019Fernandes, 2021Fulton}} indicate that the peak of the extrasolar giant planet distribution lies at ${\sim}2{-}3\,\rm{au}$, which coincides well with the water frost lines for solar type stars where planet formation is thought to be more efficient. {Theoretical studies \citep[e.g.,][]{2019Frelikh} also point to an increased abundance at this orbital separation.} 

Due to the fundamental limiting resolution of $8{-}10\,\rm{m}$ telescopes at near-infrared wavelengths, recent DI searches \citep[][]{Vigan2021A&A} can barely reach frost line separations of ${\sim}3\,\rm{au}$. 
Only $\sim$20 stars from the first targets of the \textit{SHINE} {coronagraphic} survey \citep[][consisting of 150 stars]{2021Desidera,2021Langlois,Vigan2021A&A} using \textit{VLT}/\textit{SPHERE} had the combination of youth and proximity to reach sensitivities of ${\sim}10\,\rm{M\textsubscript{Jup}}$ exoplanets at ${\sim}3\,\rm{au}$. And less than $\sim$5 stars from this survey allowed sensitivities to ${\sim}3\,\rm{M\textsubscript{Jup}}$ exoplanets at ${\sim}3\,\rm{au}$. {This orbital region for nearby stars has been recently accessed on rare occasions, but only via optical interferometry using long-baselines \citep[e.g.,][]{2020Nowak,2022HinkleyA}. But this orbital region is expected to remain largely out-of-reach for ground-based 8-10m telescopes.}  Even with \textit{JWST} \citep{2006Gardner}, the {Rayleigh} diffraction limit (${\sim}1.22$ $\lambda$/D) at wavelength ${\sim}4.5\,\rm{\mu m}$ only allows imaging of companions at ${\sim}9$\,\textendash18\,au for stars within ${\sim}50\,$\textendash$100\,\rm{pc}$. In practice, attaining even this resolution is challenging due to the presence of residual scattered starlight not suppressed by the coronagraph,  {as well as the coronagraphic inner working angle (IWA) itself. In the case of the Near Infrared Camera \citep[][\textit{NIRCam} hereon]{NIRCam} operating at ${\sim}4.6\,\rm{\mu m}$ (with the \textit{MASK430R} round coronagraph), the IWA is 0.87" (corresponding to orbital separations of ${\sim}40\,\rm{au}$ for stars at ${\sim}50\,\rm{pc}$).}
{Hence to image companions orbiting near the frost-line separations for nearby stars, an additional technique is needed to provide sensitivity at small angular separations.}

Aperture Masking Interferometry \citep[`AMI' hereon, ][]{1986Baldwin,1987Haniff,1988Readhead} achieves just this. This technique involves using an opaque mask with {a collection of strategically placed holes}, arranged in a way such that the baseline between any two holes samples a unique spatial frequency in the pupil plane. This brings the IWA down to {${\sim}0.5\lambda/D$} and has been successfully used along with Adaptive Optics (AO) from ground-based observatories \citep[e.g.][]{2000Tuthill,2006lloyd,2007Monnier,2008Woodruff,hci11, 2015Hinkley}. 
For the first time, \textit{JWST} {is executing} this on a space telescope (see \S\ref{sec:NIRISSSimulations}), taking advantage of the  exquisite sensitivity of the Near Infrared Imager and Slitless Spectrograph {\citep[][\textit{NIRISS} hereon]{DoyonNIRISS} instrument}. 
{In this work we show} that \textit{JWST} operating in the \textit{NIRISS}/\textit{AMI} mode will possess the combination of angular resolution, sensitivity and contrast to be able to access planetary mass companions at water frost-line separations around {carefully selected} nearby stars. {This capability of \textit{JWST} presents the opportunity to detect and characterize a much greater number of extrasolar giant planets and thereby constrain their early thermal histories.}

In Section \ref{sec:SampleSelection} we review the selected sample of stars for this study {composed of high-probability members of nearby young stellar associations}, followed by the simulations we used for our predictions in section \ref{sec:NIRISSSimulations}. In section \ref{sec:DetProb} we describe the conversion of these simulations to mass sensitivity limits and then subsequently to detection probabilities. In section \ref{sec:yeild_calculation} we describe {our calculation of the detection yield of planetary mass companions for these synthetic observations}. Our main results are discussed in section \ref{sec:Discussion}, and we summarise our conclusions in section \ref{sec:Conclusions}.
 

\begin{figure}
	\includegraphics[width=\columnwidth]{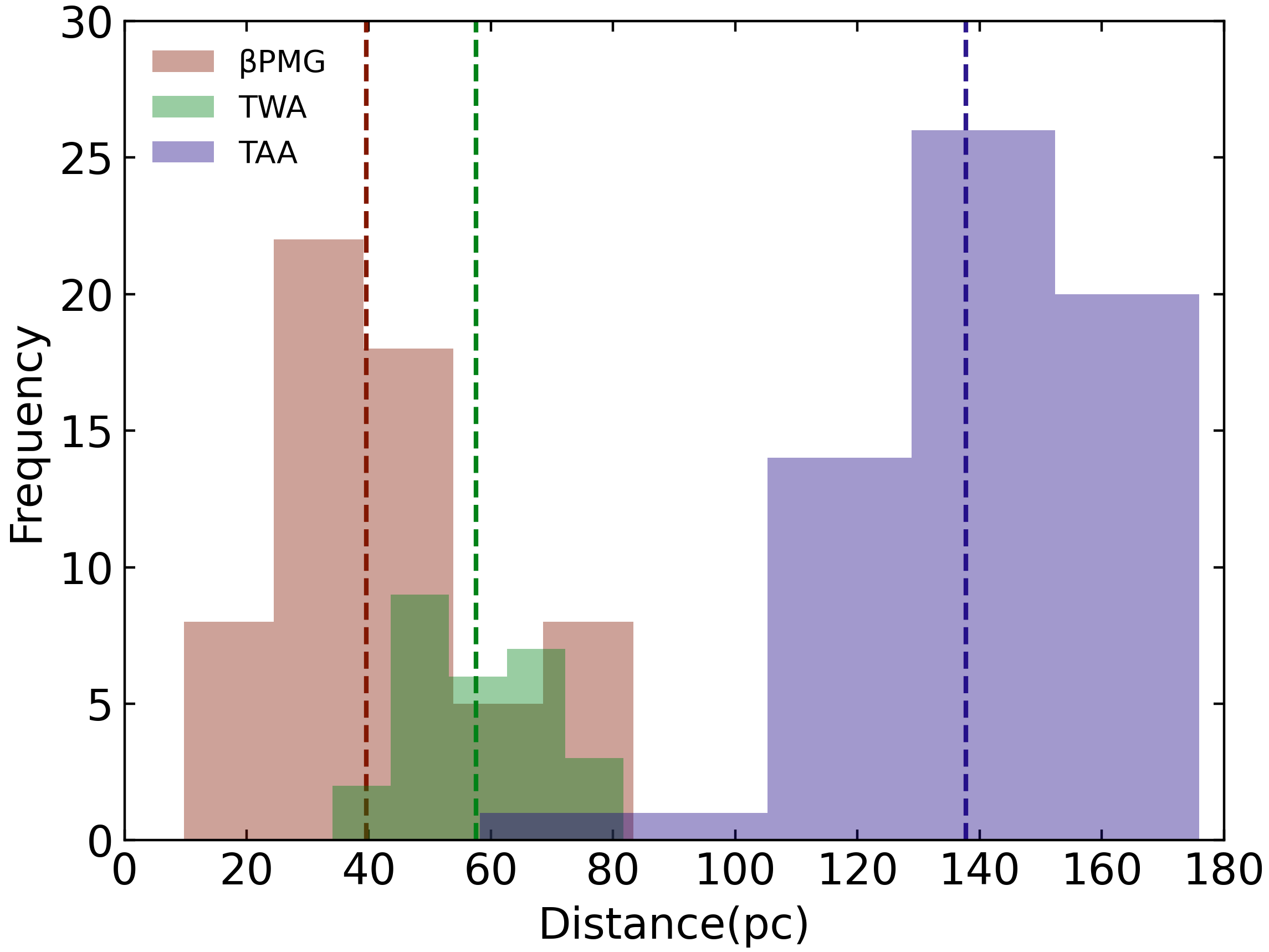}
    \caption{A histogram showing stellar distances for the samples of $\beta$Pic, TWA and TAA depicted using the colours pink, blue and green respectively. The median distance value of each sample is depicted with a dashed line using the same colour scheme.}
    \label{fig:DistanceHist}
\end{figure}

\begin{figure*}
	\includegraphics[scale=0.7]{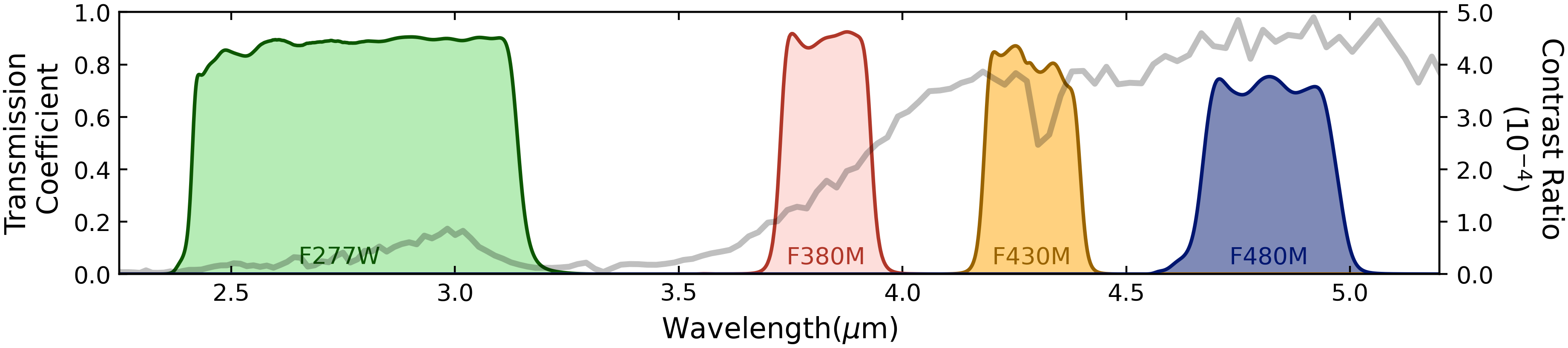}
    \caption{Transmission curves of \textit{JWST}/NIRISS filters compatible with AMI. The filters are F277W, F380M, F430M and F480M. The grey curve shows the wavelength dependent contrast ratio between a solar type star and a planet with effective temperature of $1000\,\rm{K}$ and log(g) of 4 \protect{\citep[using the equilibrium case grids from][]{PhillipsPaper}}. The filters F430M and F480M are used in this study since exoplanets have enhanced luminosity at these wavelengths thereby reducing the overall brightness difference relative to the host stars.}
    \label{fig:TransmissionCurves}
\end{figure*}

\section{Sample Selection}
\label{sec:SampleSelection}
For the purposes of this work, a part of the sample of nearby stars selected was the same as in \cite{Carter2021}, which was comprised of the stars in the $\beta$~Pictoris Moving Group \citep[][$\beta$Pic hereon]{1997Sci...277...67K} and TW Hydrae Association \citep[][TWA hereon]{2001ApJ...562L..87Z}. Although many moving groups consist of stars which provide a combination of age and distance suitable for directly imaging exoplanets \citep{GagnePaper}, $\beta$Pic and TWA associations satisfy all of the following conditions, 
making them ideal collections of targets:
\begin{itemize}
\item  distances close enough to favourably probe the innermost architectures of planetary systems through direct imaging \citep{GagnePaper}
\item  ages old enough that planetary formation processes have largely ended due to disk clearing \citep{HaischPaper}
\item ages young enough that any potentially formed planet has retained a significant amount of heat from its initial gravitational contraction and are therefore {will have a luminosity enhanced by orders of magnitude relative to field stars} \citep{BaraffePaper,PhillipsPaper}
\end{itemize}


For calculations involving the estimation of mass contrast limits using evolutionary models (see section \ref{sec:MassSensitivityLimits} for details), the ages used for the stars in the samples of $\beta$Pic and TWA were $24$$\pm3\,\rm{Myr}$ and $10$$\pm3\,\rm{Myr}$ respectively \citep{2014Malo,2015Bell}. 

In addition to $\beta$Pic and TWA, a list of confirmed members of stars in the $1{-}2\,\rm{Myr}$ Taurus-Auriga Association \citep[][TAA hereon]{taurusPaper} taken from \cite{2017Kraus} were also used in this analysis. In addition to the significantly younger age of the selected stars, the TAA sample has the advantage of the targets being highly localized on the sky compared to either the TWA or $\beta$Pic moving groups, which could potentially lead to an enhanced efficiency for a future survey (see \S\ref{sec:Conclusions}).  {There is evidence that the overall stellar population in TAA is comprised of a younger subpopulation of stars with ages of $1{-}2\,\rm{Myr}$ and an older subpopulation with ages as old as ${\sim}40\,\rm{Myr}$ \citep{2017Kraus}}. To {address} this issue, the age of each member was calculated using a methodology similar to the one used in \cite{sq2021}, which is detailed in section \ref{sec:stellar_mass}.
{Those stars for which our analysis returned a calculated age ${<}4\,\rm{Myr}$ were assigned an age of $2\,\rm{Myr}$ to match the age of $1{-}2\,\rm{Myr}$  that has been well established in previous works on the age of the underlying younger population of stars in TAA. This exercise eradicates any bias in our results from the older population (${\sim}40\,\rm{Myr}$), and assigning a single age to this younger population ensures that our analysis will be consistent with the single-age methodology we use for the $\beta$Pic and TWA samples}.


Our final sample contained 150 stars, comprised of 61, 27, and 62 stars from  $\beta$Pic, TWA and TAA respectively (see Figure \ref{fig:DistanceHist}). After selecting the stars for this study, {the synthetic} contrast curves {measuring the sensitivity of the \textit{JWST/NIRISS/AMI} mode} in terms of magnitude, were calculated using existing simulations, as detailed below.

\section{\textit{NIRISS} AMI simulations}
\label{sec:NIRISSSimulations}
The \textit{JWST/NIRISS} instrument provides high-contrast interferometric imaging using a  non-redundant mask \citep[][]{SivaramkrishnanNRM}, which turns a filled aperture into an interferometric array.  This mode offers the possibility of pushing the planet detection parameter space to well within $\lambda/D$. 
This mask is an opaque element with 7 hexagonal apertures. These hexagons when projected onto the \textit{JWST} primary mirror,{ have an incircle diameter} of approximately $0.8\,\rm{m}$ \citep[][]{2012Sivaramkrishnan,greebaum2015}. 
Using it in conjunction with the \textit{NIRISS} filters (F277W, F380M, F430M, and F480M, see Figure \ref{fig:TransmissionCurves}), this observing mode can probe objects with the highest angular resolution compared to any other mode on \textit{JWST} \citep{Artigau2014}, and offers the possibility of observing faint targets that would otherwise be inaccessible to ground-based AO facilities. 

{To detect companions to the stars in the groups of $\beta$Pic, TWA, and TAA, a desired contrast should be chosen to optimise the potential of making such a discovery. This should then be followed by choosing a particular technique such as AMI or KP interferometry \citep[e.g.,][``KP'' hereon]{2010Martinache} to execute this. Due to the comparable contrast performance between AMI and KP in the brightness range of the stars considered in this paper, for our analysis we have chosen to utilize the AMI contrast curves as a representative interferometric contrast that can be achieved with \textit{JWST}. But the actual technique can be chosen later when planning the observations, depending on the brightness of the host star (see \S\ref{ssec:kernelPhase} for more details).} To simulate the performance of this mode for particular stars in the moving groups of $\beta$Pic, TWA, and TAA, the results from \cite{StephJatis} were used which are discussed below. 

\subsection{Calculating NIRISS AMI contrast curves for a star of any given magnitude}
\label{ssec:ContrastCurves}
\cite{StephJatis} simulated {\textit{NIRISS}/AMI observations}, which were computed using the engine \texttt{Pandeia} \citep{PandeiaReference} and the software \texttt {WebbPSF} \citep{WebbPSFReference}, both of which are developed using the \textsc{Python} language. {\cite{StephJatis} also simulated \textit{NIRCam} KP observations, which does slightly outperform the AMI observations in certain cases (see \S\ref{ssec:kernelPhase} for more details). In this study we choose to instead focus on the \textit{NIRISS/AMI} results, which were used to obtain the contrast curves for our study, the methodology for which is explained below.}


\begin{figure}
	\includegraphics[width=\columnwidth]{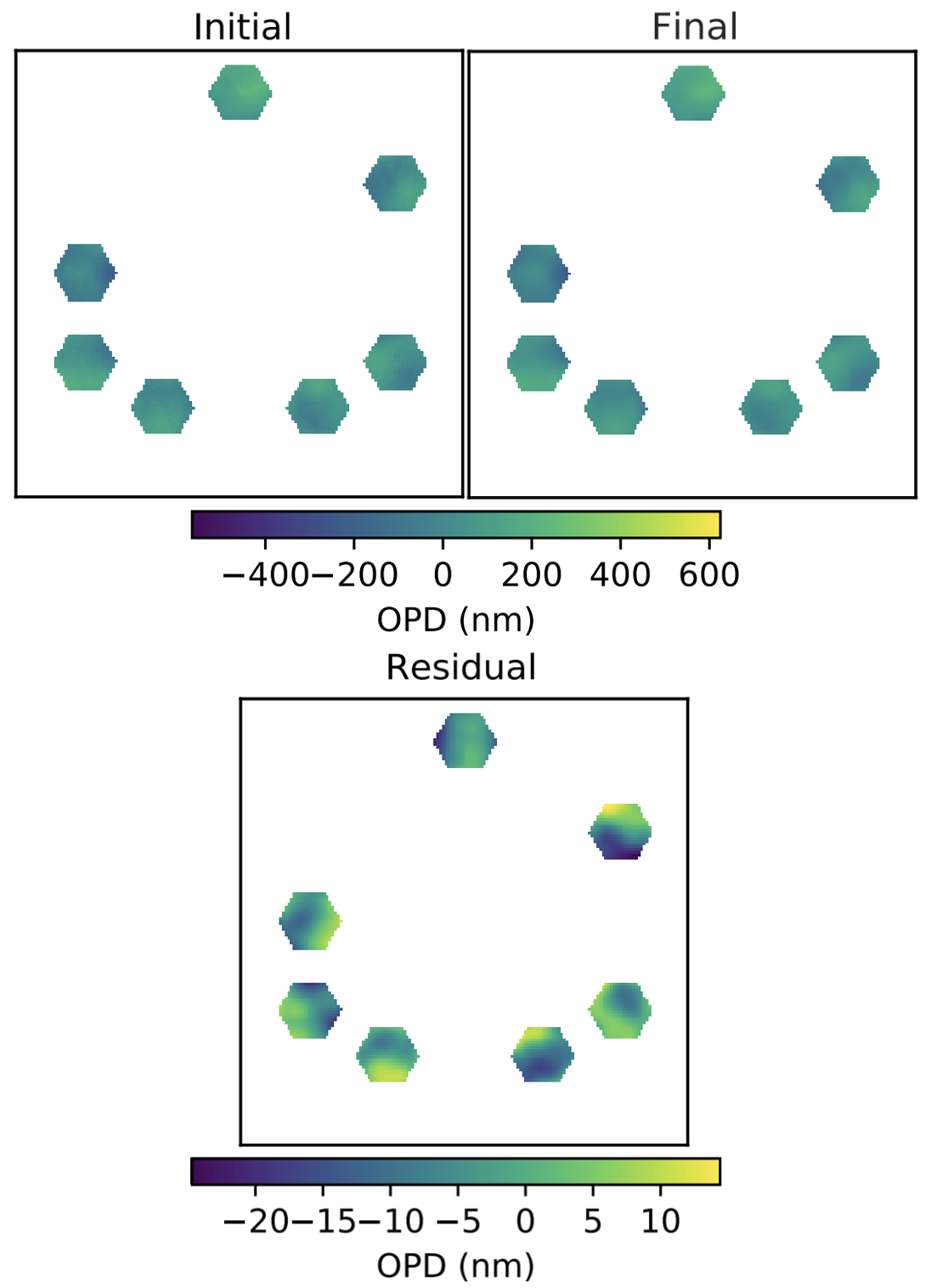}
    \caption{, A single simulation of the initial, final and residual Optical Path Difference  map for \textit{JWST}/NIRISS, recreated from \protect\cite{StephJatis}.}
    \label{fig:Initial,Final,RwsidualOPD}
\end{figure}

\begin{figure}
	\includegraphics[width=\columnwidth]{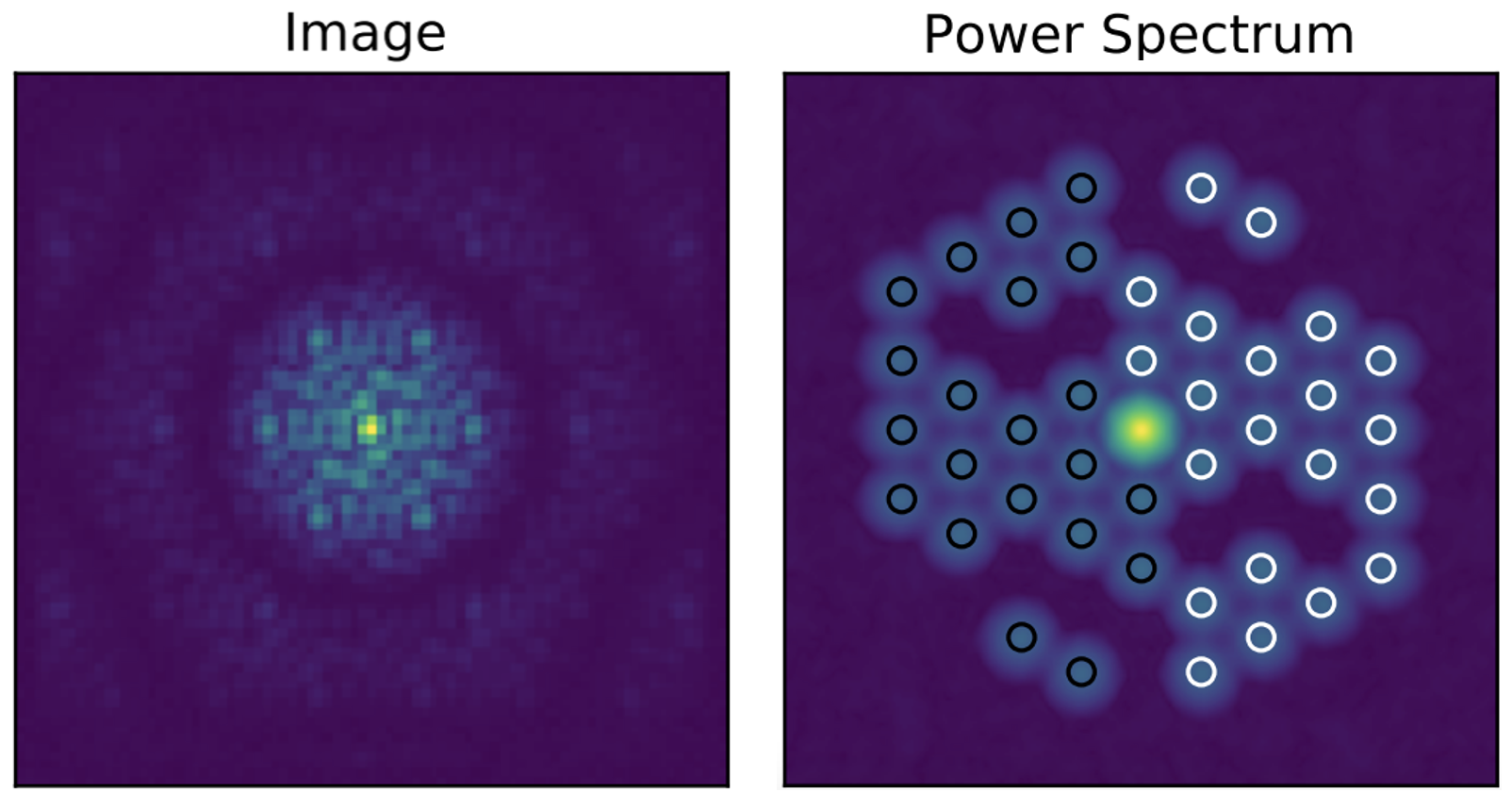}
    \caption{Simulated NIRISS interferograms (left) and power spectra (right) for a star, recreated from \protect\cite{StephJatis}}
    \label{fig:NIRISSSimulationsImage}
\end{figure}

Since exoplanets are relatively bright in the ${\sim}4{-}5\,\rm{\mu m}$ part of the spectrum when compared to their host stars (see Figure \ref{fig:TransmissionCurves}), the simulations were carried out in filters centred on these wavelengths that can be used with the \textit{NIRISS/AMI} mode, namely F430M and F480M. These simulated observations were composed of two pairs of target - Point Spread Function (PSF hereon) calibrator visits taken at different telescope roll angles $45^{\circ}$ apart, under the assumption that the length of each visit was $1.5\,\rm{hours}$, and the total observation time was $6\,\rm{hours}$. Although the maximum roll angle for \textit{JWST} at a given time is ${\sim}15^{\circ}$, the $45^{\circ}$ apart simulated visits do not significantly change the contrast curves, since the fourier coverage of the \textit{NIRISS} mask is relatively uniform. Thus, the two visits at $0^{\circ}$ and $15^{\circ}$ respectively should have a similar effect on the ability to recover companions as the two at $0^{\circ}$ and $45^{\circ}$, especially since reference PSFs are used (rather than angular differential imaging).
As each visit of a \textit{JWST} observation is split into sets of integrations, which are in turn comprised of a number of groups \citep{PandExoReference}, the maximum number of groups ($n_g$) was calculated that can be used in a single integration without saturation for a star of a given magnitude. Then a visit is constructed with the maximum number of such integrations ($n_i$) that can be acquired in $1.5\,\rm{hours}$, noting that each integration comes with a readout overhead of $0.0745\,\rm{s}$ \citep[in a \texttt{sub80} subarray,][]{Jdox}. 
When the remaining time after $n_i$ integrations allowed for more than a single group, an additional integration containing $n_{g,r}$ groups was added. A list containing the values of $n_g$, $n_i$ and $n_{g,r}$ for each calculated magnitude in the F430M and F480M is provided in Table \ref{table:Simulations} in the appendix, which is recreated from similar tables in \cite{StephJatis}. Using the image for the entire visit, science target and calibrator frames were generated using different optical path difference (OPD) maps from \texttt {WebbPSF}. This was followed by fitting a hexike \citep[hexagonal version of zernike,][]{upton2004} basis to each mirror segment with 100 coefficients. Finally, each hexike coefficient ($C_{n}$) is evolved by a factor drawn from a one-mean uniform distribution of width $2h$ tuned to result in a root mean square residual wave front error of ${\sim}10\,\rm{nm}$ with OPD evolution (see Figure \ref{fig:Initial,Final,RwsidualOPD}), shown in the following equation,


\begin{equation}
    C_{n,seg,cal}=\mathrm{Unif}(1-h,1+h) C_{n,seg,targ}
    \label{eq:OPDNIRISS}       
\end{equation}

where $h=0.2$ (for NIRISS) and the calculation is consistent with thermal evolution expected over hour long timescales for \textit{JWST}. Using these, the simulated images were computed (see Figure \ref{fig:NIRISSSimulationsImage}) from which the $5\sigma$ contrast curves were extracted. 

\begin{figure}
	\includegraphics[width=\columnwidth]{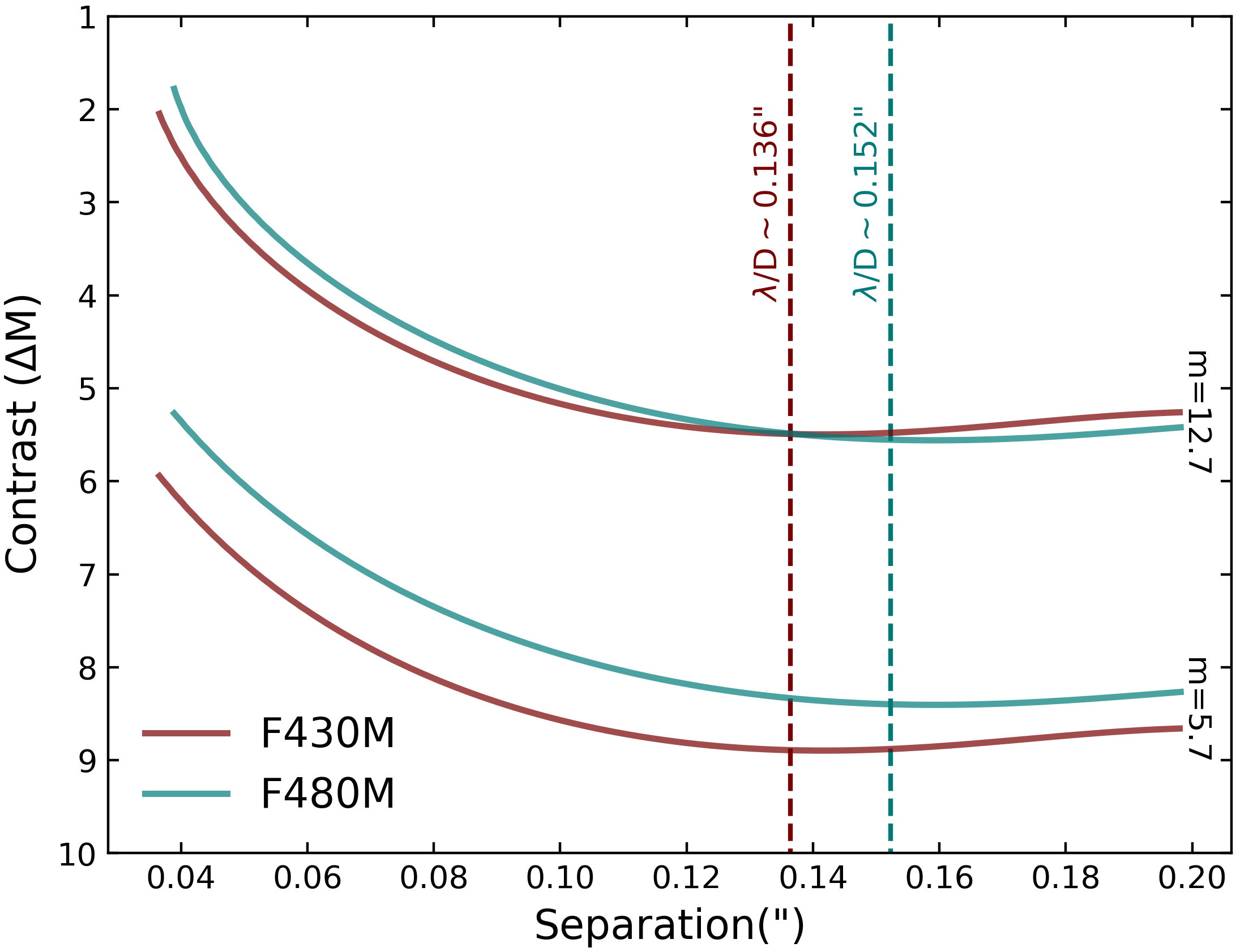}
    \caption{The available 5$\sigma$ contrast curves from the simulated images, with the lowest and the highest apparent magnitude ($m$) values of 5.7 and 12.7 respectively. The F430M and F480M \textit{JWST}/\textit{NIRISS} filters are shown in maroon and blue respectively. The classical diffraction limit is shown with dashed lines for the central wavelength of each filter with the same colour scheme.}
    \label{fig:ContrastCurvesForFilters}
\end{figure}

\begin{figure*}
	\includegraphics[scale=0.5]{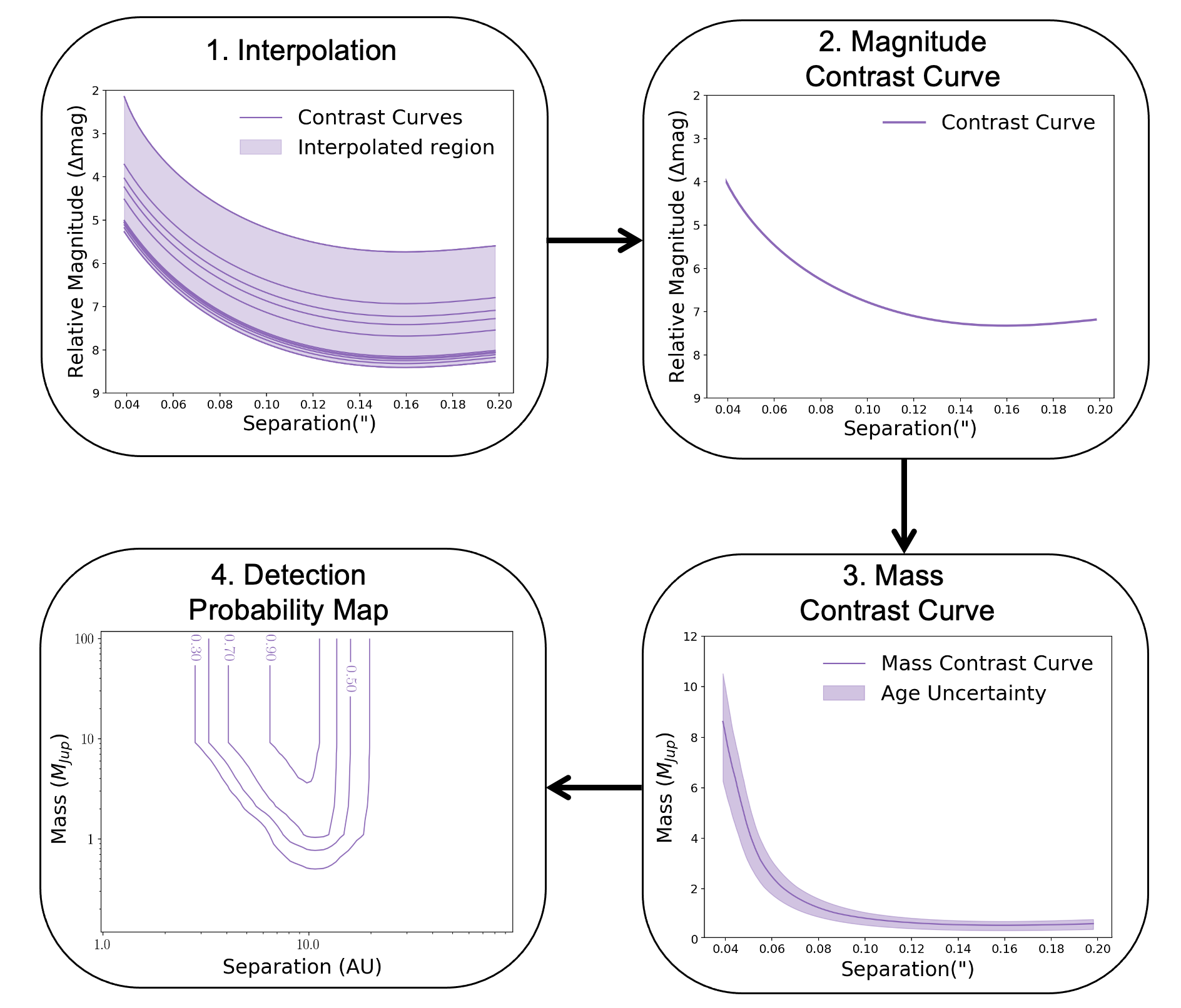}
    \caption{A summary of the methodology in this study: We start of by interpolating across the available contrast curves (1) from \protect\cite{StephJatis} and then proceed to calculate the filter specific contrast curves for all the members of the sample with the filter specific contrast (2). We then convert all these contrast curves to mass sensitivity limits using the models described in  \protect\cite{PhillipsPaper} using the ages of the moving groups (3) with an age uncertainty (we only however use the central age value going forward). Finally we calculate the detection probability maps (4) from these mass sensitivity limits for all the stars using \texttt{Exo-DMC} \protect\citep{ExoDMC}.} 
\label{fig:MethodFC}
\end{figure*}

The contrast curves were hence available for stellar apparent magnitude values ranging from 5.7 to 12.7 with increments of 0.1 (see Figure \ref{fig:ContrastCurvesForFilters}). This discrete parameter space was made continuous by interpolating across relative magnitude values for each of the apparent magnitudes values (see Figure \ref{fig:MethodFC}). This allowed us to compute the filter-specific contrast curve of all the stars in our sample given their apparent magnitude value. These filter-specific stellar apparent magnitude values for stars were calculated using stellar  isochronal models and is detailed in the following section. 

\subsection{Calculating magnitudes of stars in particular filters} 
\label{ssec:FilterMagnitudeCalculation}
Apparent stellar magnitudes in the \textit{JWST/NIRISS/AMI} filters of F480M and F430M for the stars in our samples of $\beta$Pic and TWA were calculated following a similar methodology as that described in \citet{Carter2021}, which is briefly outlined here for clarity. To begin, the effective temperature~($T_\mathrm{eff}$) and log($g$) was estimated for each star in the sample by matching their \textit{Gaia} \citep{Gaia16, Gaia18} $B{-}R$ colours to the theoretical stellar isochrones covering $0.07{-}1.4\,\rm{M_{\odot}}$ \citep{Bara15} and $0.8{-}120\,\rm{M_{\odot}}$ \citep{Haem19}. A spectral energy distribution (SED) for each star was then determined by matching its estimated $T_\mathrm{eff}$ and log($g$) to interpolated solar metallicity spectra obtained from \citet{Bara15}. 
These spectra were then normalised using the respective magnitudes of the corresponding star's \textit{WISE} $W2$ ($m\textsubscript{W2}$) bandpass \citep{Wrig10, Cutr13}. Finally, the apparent magnitudes for each star in the \textit{NIRISS/AMI} filters were calculated using the \texttt{pysynphot} \textsc{Python} package \citep{Pysy2013}. 

For the stars in TAA, given their young age, only the isochrones from 
\cite{Bara15} were used since {these models have a lower age limit of $1\,\rm{Myr}$}. However, some of the stars in the sample do not have \textit{Gaia}  $B$-$R$ colour magnitude values in the domain of these isochrones. 
To solve this, the filter specific magnitudes ($m\textsubscript{F430M}$ and $m\textsubscript{F480M}$) for the stars which did have $B$-$R$ colour magnitude values in the domain were first calculated using the same method as the stars in $\beta$Pic and TWA. The magnitudes of the remaining stars were calculated from an interpolation of $m\textsubscript{F430M}$ versus $m\textsubscript{W2}$ magnitudes and $m\textsubscript{F430M}$ vs. $m\textsubscript{W2}$ magnitudes separately, by reading off their respective $m\textsubscript{W2}$ magnitudes. 

Using these apparent magnitudes, the contrast curves (representing the achieved sensitivity to faint companions, measured in  magnitudes fainter than the host star) were computed for each star in the sample using the generated interpolated parameter space as discussed in section \ref{ssec:ContrastCurves}. These contrast curves were then converted into values in terms of mass using the evolutionary models described in \cite{PhillipsPaper}, as detailed in the following section (see Figure \ref{fig:MethodFC}).

\section{Calculation of Detection Probabilities}
\label{sec:DetProb}
In this section we describe our calculations of the probability of detecting substellar companions as a function of mass and orbital separation for each of the targets within our sample.  

\subsection{Mass sensitivity limits}
\label{sec:MassSensitivityLimits}
\texttt{ATMO} 2020 is a set of 1D radiative-convective equilibrium cloudless models describing the atmosphere and evolution of cool brown dwarfs and self-luminous giant exoplanets \citep{PhillipsPaper}, spanning the mass range of  ${\sim}0.5\,\rm{M\textsubscript{Jup}}{-}75\,\rm{M\textsubscript{Jup}}$.
This set of models was used to convert the obtained contrast curves to mass sensitivity limits at given separations. \texttt{ATMO} offers three different sets of evolutionary models: one at
chemical equilibrium, and the other two at chemical disequilibrium assuming different strengths of vertical mixing. We keep our calculations and results limited to the case of equilibrium models, {since this case provides the baseline scenario of planetary atmospheric conditions, eliminating more complex considerations related to atmospheric dynamics, such as vertical atmospheric mixing \citep[][]{bmk11, 2013Konopacky}. Although some planetary mass companions do show signs of disequilibrium chemistry, for simplicity this study does not take disequilibrium models into consideration.}
Using these mass sensitivity limits hence calculated for each star in the sample, the detection probabilities of were calculated (see Figure \ref{fig:MethodFC}). 


\subsection{Mapping the probability of detecting companions}
\label{sec:dmc_intro}

The Exoplanet Detection Map Calculator \citep[\href{https://ascl.net/2010.008}{\texttt{Exo-DMC}},][]{ExoDMC} was used to estimate detection probability maps of companions for the stars in the sample. This \textsc{Python} language tool is an adaptation of the previously existing code \texttt{MESS} \citep[Multi-purpose Exoplanet Simulation System,][]{MESS} and uses a Monte Carlo approach to compare the instrument detection limits with a simulated, synthetic population of planets {with varying orbital geometries} around a given star to estimate the probability of detection of a companion of a given mass and semi-major axis. This information is then summarised in a detection probability map. 

For all the stars in the sample (members of $\beta$Pic, TWA, and TAA), \texttt{Exo-DMC} was used to produce a population of synthetic companions with masses and semi-major axes from $0.1\,\rm{M\textsubscript{Jup}}$  to $100\,\rm{M\textsubscript{Jup}}$ and $1\,\rm{au}$ to $1000\,\rm{au}$ respectively. For each point in the mass/semi-major axis grid, \texttt{Exo-DMC} generates a fixed number of sets of orbital parameters. As discussed in \cite{Bonavita2013}, all the orbital parameters are uniformly distributed except for the eccentricity, which is generated using a Gaussian distribution with $\mu =0$ and $\sigma = 0.3$, following the approach by \cite{Hogg2010}. This approach takes into account the effects of projection when estimating the detection probability using the calculated mass sensitivity contrasts (see section \ref{sec:MassSensitivityLimits}) by estimating the projected separations corresponding to each orbital set for all the values of the semi-major axis in the grid \citep[see][for a detailed description of the method used for the projection]{MESS}. The detection probability of each synthetic companion is therefore calculated as its probability to truly be in the instrument field of view and therefore to be detected, if the value of the mass is higher than the contrast limit.

\texttt{Exo-DMC}'s basic setup uses a flat logarithmic distribution for both mass and semi-major axis. However, there is a high level of flexibility in terms of possible assumptions on the synthetic planet population to be used for the determination of the detection probability. To fully understand this feature, one needs to keep in mind that \texttt{Exo-DMC}'s detection probability is in fact made up of two terms: the probability of the companion of a given mass and semi-major axis to exist; and the probability of it being in the field of view and above the detection threshold set by the calculated detection limits, as described in sections \ref{sec:NIRISSSimulations} and \ref{sec:MassSensitivityLimits}. In the default setup, the standard assumption is that each companion in the grid has the same probability to exist. 
Changing the assumption on the companion parameter distribution does not change the shape of the detection contours, so the sensitivity remains the same, but the chances of a companion of a given mass to actually be there become unequal across the grid.

Finally, regardless of the parameter distribution used, the underlying assumption is that each target can lead to only one detection. Therefore to use the output from the \texttt{Exo-DMC} runs, to estimate the overall survey yield one needs to apply an appropriate normalisation factor ($C_0$), which is usually defined so that the expected number of detections in a given mass and semi-major axis range reflects the observed value in that same range (see section \ref{ssec:YeildCalculation}).

The probability maps hence generated for each star in the sample were then averaged in a cumulative manner after ranking them by lower values of mass and semi-major axis, as discussed in the following section. 

\subsection{Ranking the detection probability maps}
\label{ssec:RankingBox}

Selecting the best targets from the sample to observe, to find companions around them in relatively close in separations, is the most direct path to answering crucial questions about {their initial entropies (see section \S\ref{ssec:CloseInSeparations})}. In addition to discarding the stars from a survey which have low (${\lesssim}5\%$) probabilities of companion detectability in the said parameter space, it also enables the selection of a limited number of candidates companions to target with the telescope, saving on expensive observing time. 

\begin{figure}
	\includegraphics[width=\columnwidth]{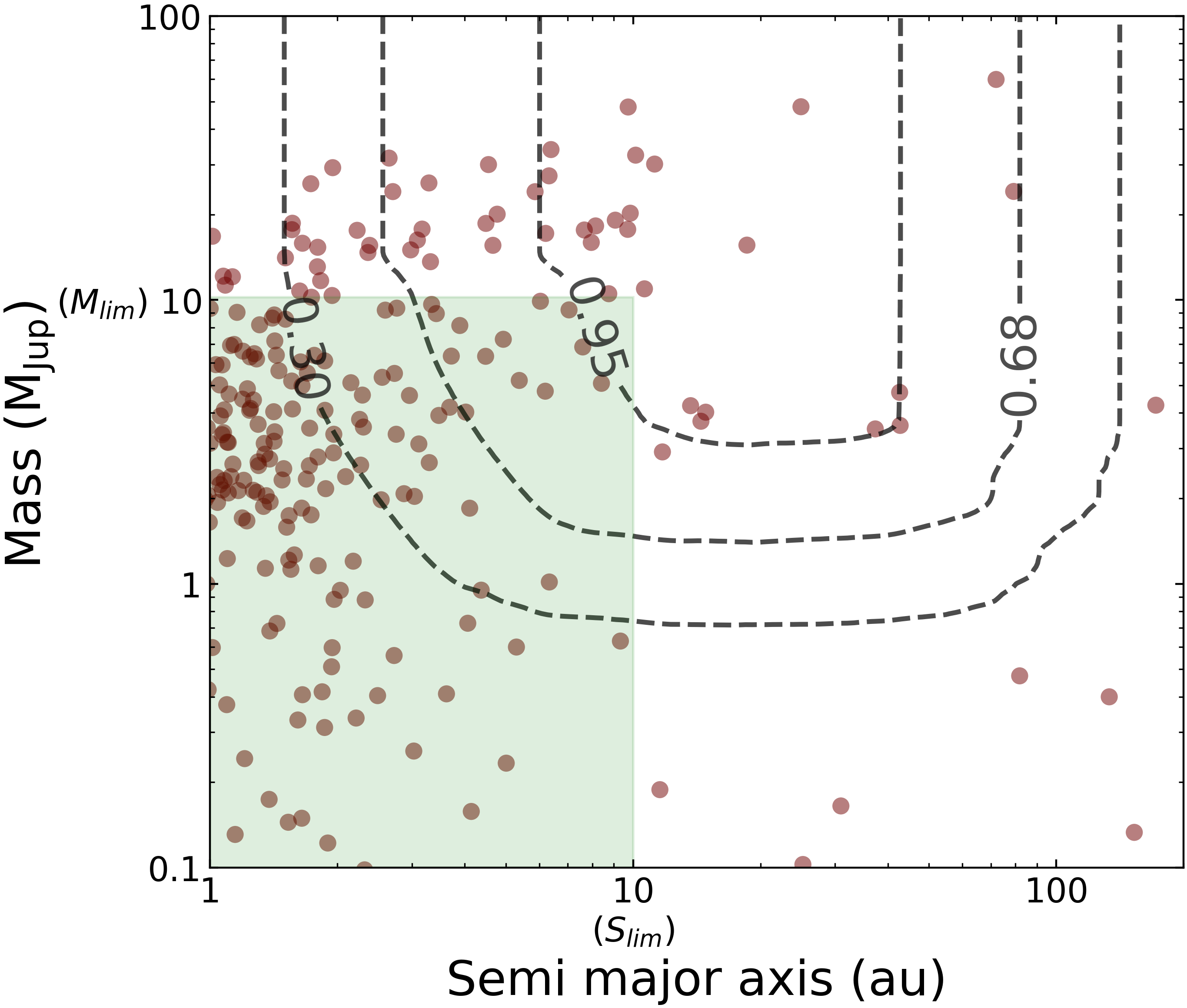}
    \caption{A plot showing the region (in green) in which the companion detection probability is summed over every grid point bounded by $M_{\rm lim}$ and $S_{\rm lim}$ to rank the stars in the sample. {The detection probability map is used as an example to show the overlap between itself and the green region, which is the average completeness map for the best 40 members in $\beta$Pic. The \protect\cite{2021Burn} synthetic planetary population is shown using brown circles, which is analogous to the \protect\cite{2021Fulton} distribution.}}
    \label{fig:RankingPlot}
\end{figure}

To rank the targets based on their potential to detect planetary mass objects at frost-line type separations, a region in the mass/separation space was first defined. This was defined as the region in detection probability maps where the mass and separation are below the values of $M_{\rm lim}$ and $S_{\rm lim}$ respectively (see Figure \ref{fig:RankingPlot}). Following this, the probability of finding a companion at each grid point was added for all the companions in this parameter space region and all the members of each sample were ranked from the highest probability to the lowest probability. Mathematically, for each member $i$  of the sample, the associated rank $\mathcal{R}_i$, is given by,

\begin{equation}
     \mathcal{R}_i =\sum^{M_{lim}}_{y=0.1}\sum^{S_{lim}}_{x=1}p_{i}(x,y)
     \label{eq:Ranking}       
\end{equation}

where $p_{i}(x,y)$ is the probability detection value at each grid point and $x$ and $y$ are the semimajor axis and the mass values respectively. The values of $M_{\rm lim}$ and $S_{\rm lim}$ for this project were selected as $10\,\rm{M\textsubscript{Jup}}$ and $10\,\rm{au}$ respectively. The semimajor axis upper limit was chosen to ensure sensitivity to frost-line separations (see section \ref{ssec:CloseInSeparations}), while the upper limit of $10\,\rm{M\textsubscript{Jup}}$ was chosen since the hot and cold-start luminosity evolution models can be more easily distinguished for more massive planetary mass companions, (${\gtrsim}10\,\rm{M\textsubscript{Jup}}$), as discussed for example in Figure 7 of \cite{2012Spiegel}, see \S\ref{ssec:CloseInSeparations} for more details. However we do not extend this upper limit beyond $10\,\rm{M\textsubscript{Jup}}$, in order to remain in the mass regime consistent with planetary mass companions. As shown in Figure \ref{fig:RankingPlot}, this region is also where there is an increased density in the synthetic planet population from \cite{2021Burn}.


The next step was to select a number of candidates from the sample based on this ranking.  
Once all the stars in each sample were ranked using equation \ref{eq:Ranking} and a list was created, the average detection probability map was calculated. This was done by taking the mean detection probability map of the $N$ best candidates, where $N\in[1,N_{tot}]$ and $N_{tot}$ is the total number of objects in each sample. Hence, $N_{tot}$ plots were created for each sample ($N=2$ was the average of the best two stars according to the ranked list (the  $\mathcal{R}_i$ value), $N=3$ had the average of the best three stars, etc.). After the ranked list was created and average detection probability map of the best $N$ stars from each sample was computed, the focus was shifted to calculating the yield (the average number of planets that would be detected with each observation) using the individual stellar masses of the members of the sample. 

\section{Estimating the planet detection yield}
\label{sec:yeild_calculation}
Ranking members in the sample using the method in section \ref{ssec:RankingBox} provides an initial prioritised list of preferred stars to target.  However, obtaining a more informative list based on the estimated planet detection yield should take into account any \textit{a priori} results on the orbital distribution of planets from previous planet detection surveys. To get such a list, we calculate the yield for the stars using the calculated detection probabilities along with the distributions obtained from previous surveys and use this value to rank them. This method also returns the number of companions that would statistically be detected with a given number of observations. 

Several works \citep[e.g.][]{Johnson:2007ApJ, Bowler:2010ApJ, Wagner:2019ApJ} provide hints that planet occurrence is likely to be influenced by the host star properties, with the stellar mass likely playing a key role. So, the estimates of the host star masses were refined, as this is expected to be a key variable in the determination of the detection yield, as described below. This is because the yield value is dependant on the stellar mass (see equation \ref{eq:pl_exp} in section \ref{ssec:YeildCalculation}). 

\subsection{Calculating stellar mass estimates}
\label{sec:stellar_mass}
In order to derive individual mass estimates for the sample of host stars, we employed the Manifold Age Determination for Young Stars {\citep[\textsc{madys,}][]{sq2021,2022Squicciarini}}. Starting from our target list, \textsc{madys} retrieved and cross-matched photometry from \textit{Gaia} EDR3 \citep{Gaia21} and 2MASS \citep{2mass}, and then applied a correction for interstellar extinction by integrating along the line of sight the 3D extinction map by \cite{leike20}; the derived values of the extinction in G band ($A_G$) 
were then used to evaluate the extinction in the chosen photometric band using a total-to-selective absorption ratio $R=3.16$ and extinction coefficients $A_\lambda$ from \cite{wang19}.

\begin{figure}
	\includegraphics[width=\columnwidth]{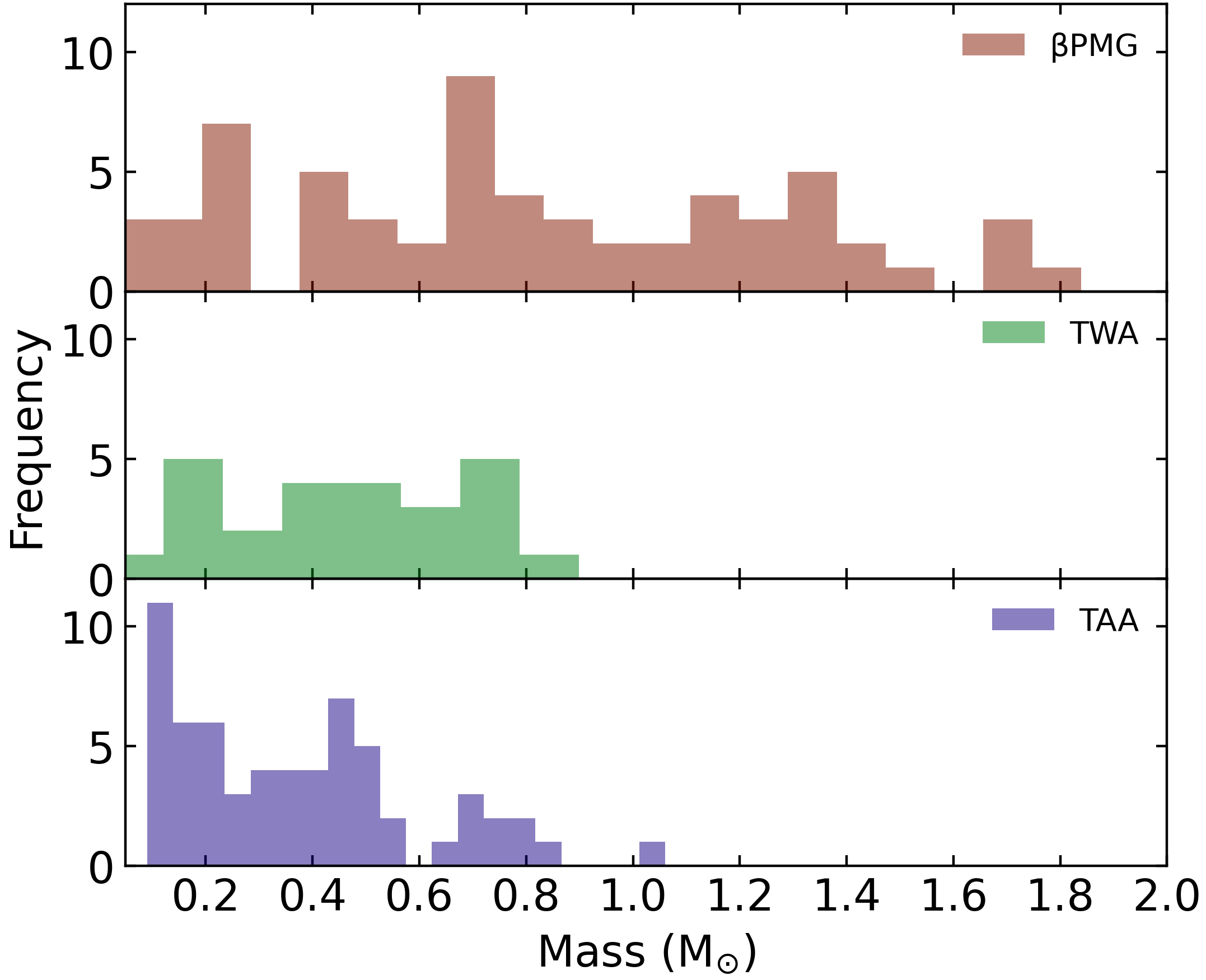}
    \caption{Histogram for the calculated masses of the members in the samples of $\beta$Pic, TWA and TAA in the first, second and third panels respectively.}
    \label{fig:MassHist}
\end{figure}

\textsc{madys} then compared, for each star, the derived absolute magnitudes with a grid of theoretical isochrones to simultaneously yield an age and mass estimate. Among several available grids, the PARSEC isochrones \citep{marigo17} were chosen, due to their large dynamical range spanning the entire stellar regime. A constant solar metallicity, appropriate for most nearby star-forming regions, was assumed \citep{dorazi11}.  Uncertainties were estimated via a Monte Carlo approach for uncertainty propagation, i.e. by replicating the computation while randomly varying, in a Gaussian fashion, photometric data according to their uncertainties. The resulting mass estimates in the form of histograms for each group are shown in Figure \ref{fig:MassHist}. {Using these mass values, the yield was calculated for all the members of the sample. This approach is more precise since it uses the photometry for each source. But since all our targets are from well known moving groups/associations (hence the ages are well constrained), we do not expect the values of the masses to have changed significantly from previous work \citep[][]{Carter2021}. So, albeit more accurate, we do not expect the change of approach for the host mass determination to have a significant impact on the final results. The absence of A stars in the TAA sample (as seen in Figure \ref{fig:MassHist}) is result of stellar evolution. In the pre-main sequence phase, stars that eventually will be earlier spectral types are still fully convective and descending down the Hayashi track. It is to be noted that there probably are at least some A stars associated with TAA. But without a clear youth indicator such as the presence of a circumstellar disk, distinguishing young early type TAA members from the field can be difficult \citep{2013Mooley}.}

\begin{figure*}
	\includegraphics[scale=0.8]{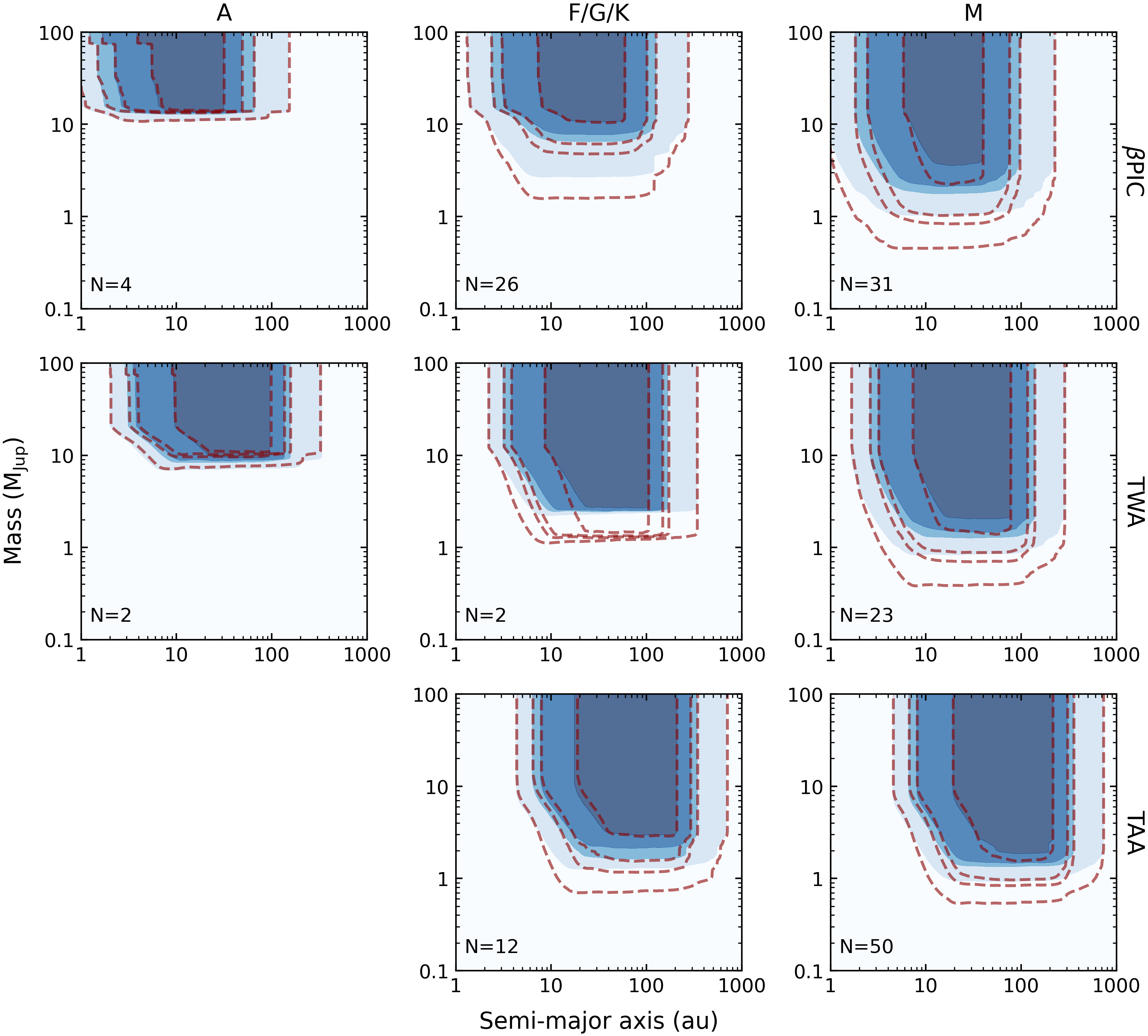}
    \caption{Average detection probability maps for the $\beta$Pic, TWA and TAA groups, separated for each of the spectral types of A (left column), F/G/K (middle column) and M (right column) in the F430M filter. {The dashed lines show the same contours for the F480M filter}. The value $N$ is the number of stars each plot is averaged over in the specific spectral type and group. The four contour lines and the four subsequently darker regions in each plot show the 10\%, 50\%, 68\%, 95\% confidences. TAA has no A stars in the sample.}
    \label{fig:SpectralClassDetProb}
\end{figure*}

\subsection{Calculating yield by fine tuning \texttt{Exo-DMC} to stellar masses}
\label{ssec:YeildCalculation}

The yield is in general evaluated by \texttt{Exo-DMC} as the convolution function 
$N_{exp} * N_{det}$, where $N_{det}$ is the {function describing the number} of detectable planets, obtained as the sum of the detection probability evaluated by the DMC {at each mass/semimajor axis grid poin}t and $N_{exp}$ is the {function describing the} expected number of planets according to the chosen set of parameter distributions, calculated as:

\begin{equation}\label{eq:pl_exp}
    N_{exp} = C_0 * 
    \int_{a_{min}}^{a_{max}}~f(a)\,da \int_{m_{min}}^{m_{max}}~f(M_p)\,dM_p 
\end{equation}

\noindent where $C_0$ is a normalisation constant which makes sure the expected frequency matches the observed one and $f(a)$ and  $f(M_p)$ are the chosen distributions for semi-major axis and mass, respectively. 

For our yield estimate, we chose to adopt two different approaches: one simply extrapolating the latest results from radial velocity (RV hereon) surveys and another from the latest DI results. {In both cases the semi-major axis follows a log-normal distribution, while the mass-ratio distribution is a power law for the planetary part and an uniform distribution for the stellar part \citep[see][for details]{Vigan2021A&A}. Below we describe both distributions in more detail.}

\begin{enumerate}

    \item {Extended Radial Velocity}\\
    The distribution is taken from \cite{2021Fulton}, and is comprised of a broken power-law for the semi-major axis and a mass distribution uniform in logarithmic scale. Although this distribution is drawn from RV data, it has been shown to agree with the DI results \citep[][]{Vigan2021A&A}, so it represent a suitable choice for our analysis. For this case the normalisation factor $C_0$ was calculated to match the results from \cite{Vigan2021A&A}, so assuming an overall frequency of 5.6\% for companions with masses between 1 and 70~$M\textsubscript{Jup}$ and separations between 5 and 300~au. \\
    
    \item { Bimodal Distribution}\\
     We also adopt the parametric model outlined in  \cite{Vigan2021A&A}. The basic assumption of this model is that the observed population is in fact made up of two components representing two different populations of substellar companions: a planet-like population and a binary star-like population.
    Each component has different parameter distributions and different normalisation factors. Also, this distribution introduces a dependence on the stellar mass, so the planet mass distribution is replaced by a mass ratio ($q=\frac{M_p}{M_*}$) distribution and the other parameters are also dependant on the primary spectral type. So Eq.~\ref{eq:pl_exp} in this case changes to: 
    
    \begin{eqnarray}\label{eq:pl_exp_mm}
    N_{exp} = C_{PL} * \int_{a_{min}}^{a_{max}}~f_{PL}(a)\,da \int_{q_{min}}^{q_{max}}~f_{PL}(q)\,dq + \\ 
    C_{BS} * \int_{a_{min}}^{a_{max}}~f_{BS}(a)\,da \int_{q_{min}}^{q_{max}}~f_{BS}(q)\,dq \nonumber
    \end{eqnarray}
    
    where the subscripts \textit{PL} and \textit{BS} refer to the planet-like and binary star-like parts of the equation respectively.  \\

    The yields hence calculated are reported in Table \ref{table:yields}. The values obtained using the extended RV distribution from \cite{2021Fulton} are lower than the ones obtained with the bimodal distribution from \cite{Vigan2021A&A} across all stellar types and groups. The yield values are essentially identical in the filters of F430M and F480M in Table \ref{table:yields}. {The highest overall yield is produced by TWA at 0.16 planets per star for the \cite{Vigan2021A&A} distribution, but only 0.07 planets per star for the \cite{2021Fulton} distribution. Meanwhile the $\beta$Pic and TAA groups have yields of $0.04{-}0.10$ planets per star.}
    

    
\end{enumerate}



\begin{figure*}
\label{table:Yields}
\centering
	\includegraphics[scale=0.8]{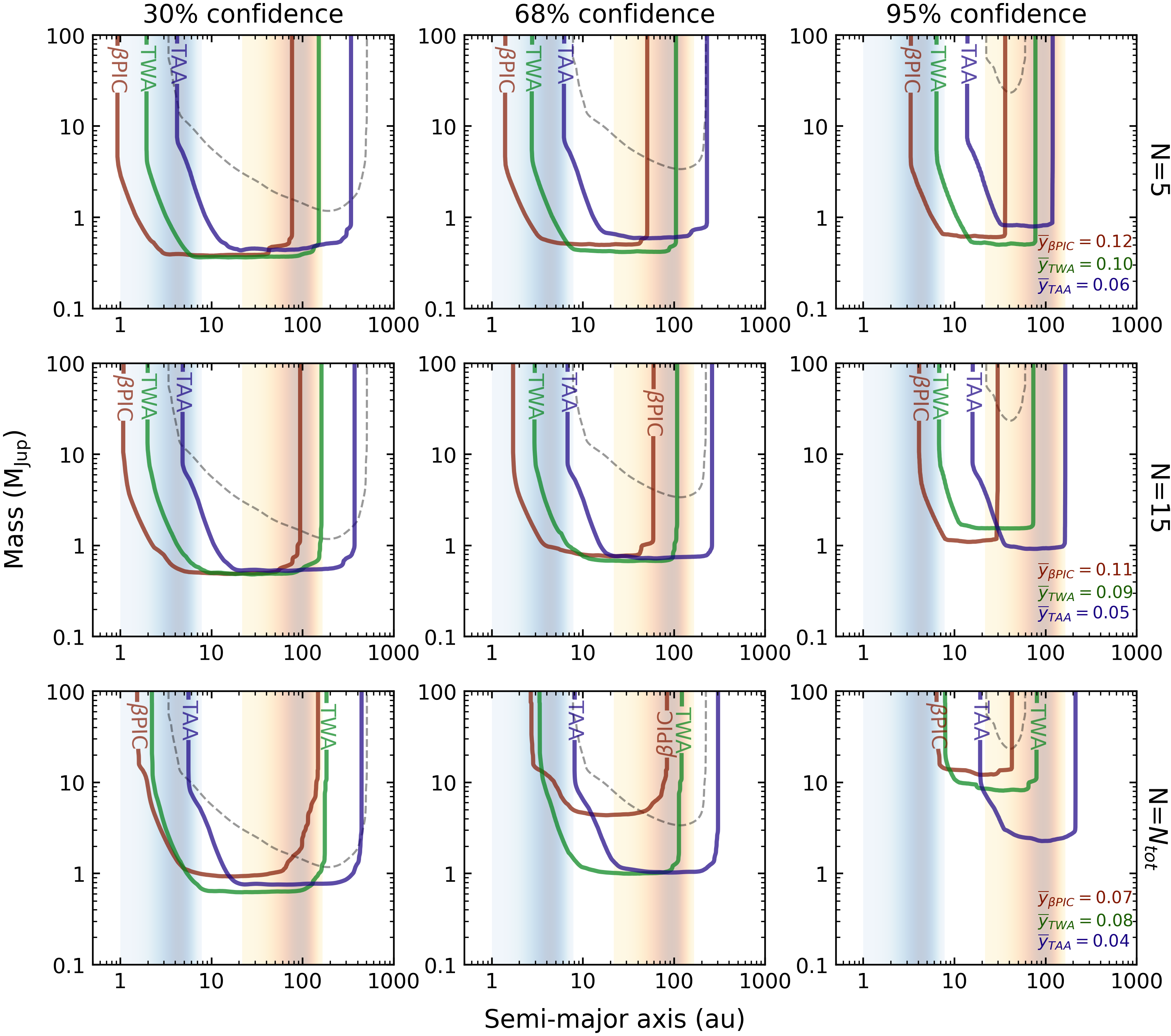}
    \caption{30\%, 68\% and 95\% confidence plots for the best $N$ stars in each of the samples of $\beta$Pic (red), TWA (green) and TAA (blue) using the F480M filter, where $N$ has been set to 5, 15 and the total number of stars ($N_{tot}$) in each sample respectively. The dashed lines are the contours for the average detection probability of the SHINE survey \protect\citep[][]{Vigan2021A&A} at the same confidence levels. The cyan and the orange gradient bands show the range of H\textsubscript{2}O and CO frost lines respectively (see section \ref{ssec:CloseInSeparations}). The third column also shows the yield value for each sample averaged over $N$ stars using the extended RV distribution \citep{2021Fulton}.}
    \label{fig:ConfidenceLevels}
\end{figure*}

\begin{table*}

\caption{Spectral type specific yield values for each of the samples of $\beta$Pic, TWA and TAA using the simulated observations in \textit{JWST}/\textit{NIRISS} filters F430M and F480M respectively. A bimodal \citep[][]{Vigan2021A&A} distribution and an extended radial velocity \citep[][]{2021Fulton} distribution were used to calculate these, as detailed in section \ref{ssec:YeildCalculation}. The mean values of yields {(number of planets per star)} for each distribution used for each stellar group are given in below the spectral type classification.}\label{table:yields}
\centering
\begin{tabular}{c|c!{\vrule}ccc!{\vrule}ccc} 
\hline
\hline
\multicolumn{2}{c|}{~}              & \multicolumn{3}{c|}{\textbf{F430M}} & \multicolumn{3}{c}{\textbf{F480M}}  \\ 
\hline
Distribution & Spectral Type & $\beta$Pic & TWA  & TAA & $\beta$Pic & TWA & TAA            \\ 
\hline
\multirow{4}{*}{\cite{Vigan2021A&A}}& A& 0.07 & 0.08 & — & 0.07 & 0.08 & —             \\ 

& F/G/K       & 0.03 & 0.04 & 0.03         & 0.04    & 0.04   & 0.04              \\ 

& M             & 0.16 & 0.17 & 0.06         & 0.16     & 0.18    & 0.06              \\ 
\cline{2-8}
 & \textbf{Mean}       & \textbf{0.10}  & \textbf{0.16} & \textbf{0.05}         & \textbf{0.10}    & \textbf{0.16}   & \textbf{0.05}              \\ 
\hline
\multirow{4}{*}{\cite{2021Fulton}} & A             & 0.04    & 0.03    & —            & 0.04    & 0.04   & —              \\ 

 & F/G/K       & 0.05    & 0.06    & 0.04            & 0.05    & 0.06   & 0.04              \\ 

& M             & 0.08    & 0.08    & 0.04            & 0.09    & 0.08   & 0.04              \\ 
\cline{2-8}
& \textbf{Mean}       & \textbf{0.07}    & \textbf{0.07}    & \textbf{0.04}            & \textbf{0.07}    & \textbf{0.08}   & \textbf{0.04}              \\

\hline
\end{tabular}

\end{table*}

\begin{figure}
\centering
	\includegraphics[width=\columnwidth]{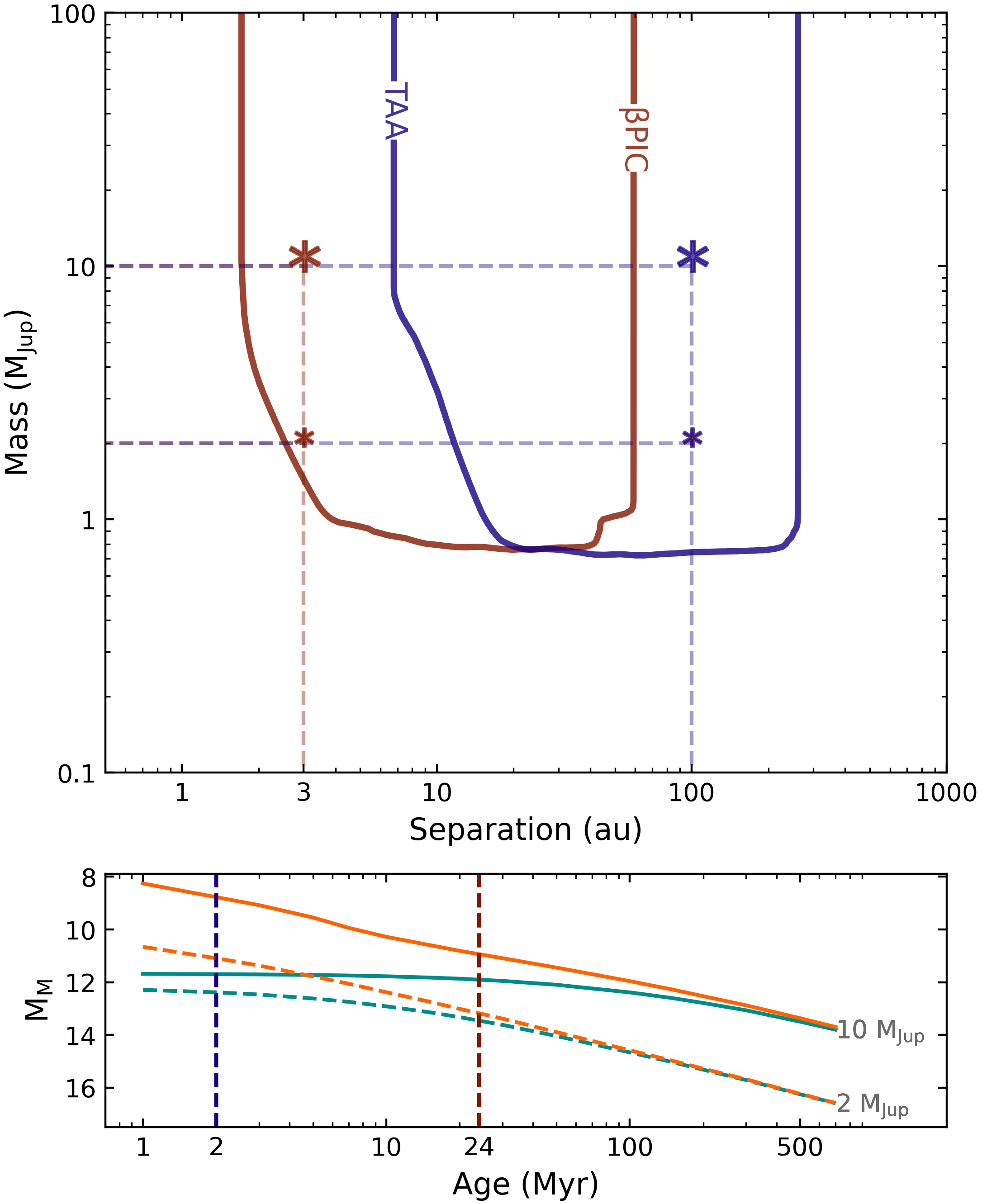}
    \caption{{Top Panel: 68\% confidence contours for $\beta$Pic and TAA for the best 15 stars. The asterisk marks are hypothetical detected planets in each moving group of mass $10\,\rm{M\textsubscript{jup}}$ and $2\,\rm{M\textsubscript{jup}}$ respectively. Bottom Panel \protect{\citep[a version recreated from][]{2012Spiegel}}: The orange and cyan solid curves show the hot (initial entropy of $13\,\rm{k_{B}/baryon}$) and cold (initial entropy of $8\,\rm{k_{B}/baryon}$) start evolution of luminosities in M band for a $10\,\rm{M\textsubscript{jup}}$ companion. The dashed lines show this for a $2\,\rm{M\textsubscript{jup}}$ companion. For these hypothetical planets, given the ages of $\beta$Pic ($\sim$$24\,\rm{Myr}$) and TAA ($\sim$$2\,\rm{Myr}$), initial entropy constraints (hot versus cold start or an intermediate value) can be probed as shown in the bottom panel. The difference in initial entropies are more pronounced for younger ages (for example it is more easily distinguishable in TAA members compared to $\beta$Pic members) and for more massive companions (for example, it is more easily distinguishable in the case of the $10\,\rm{M\textsubscript{jup}}$ companion compared to the $2\,\rm{M\textsubscript{jup}}$ companion).}}
    \label{fig:EntropyCalculation}
\end{figure}

\begin{figure*}
\centering
	\includegraphics[scale=0.8]{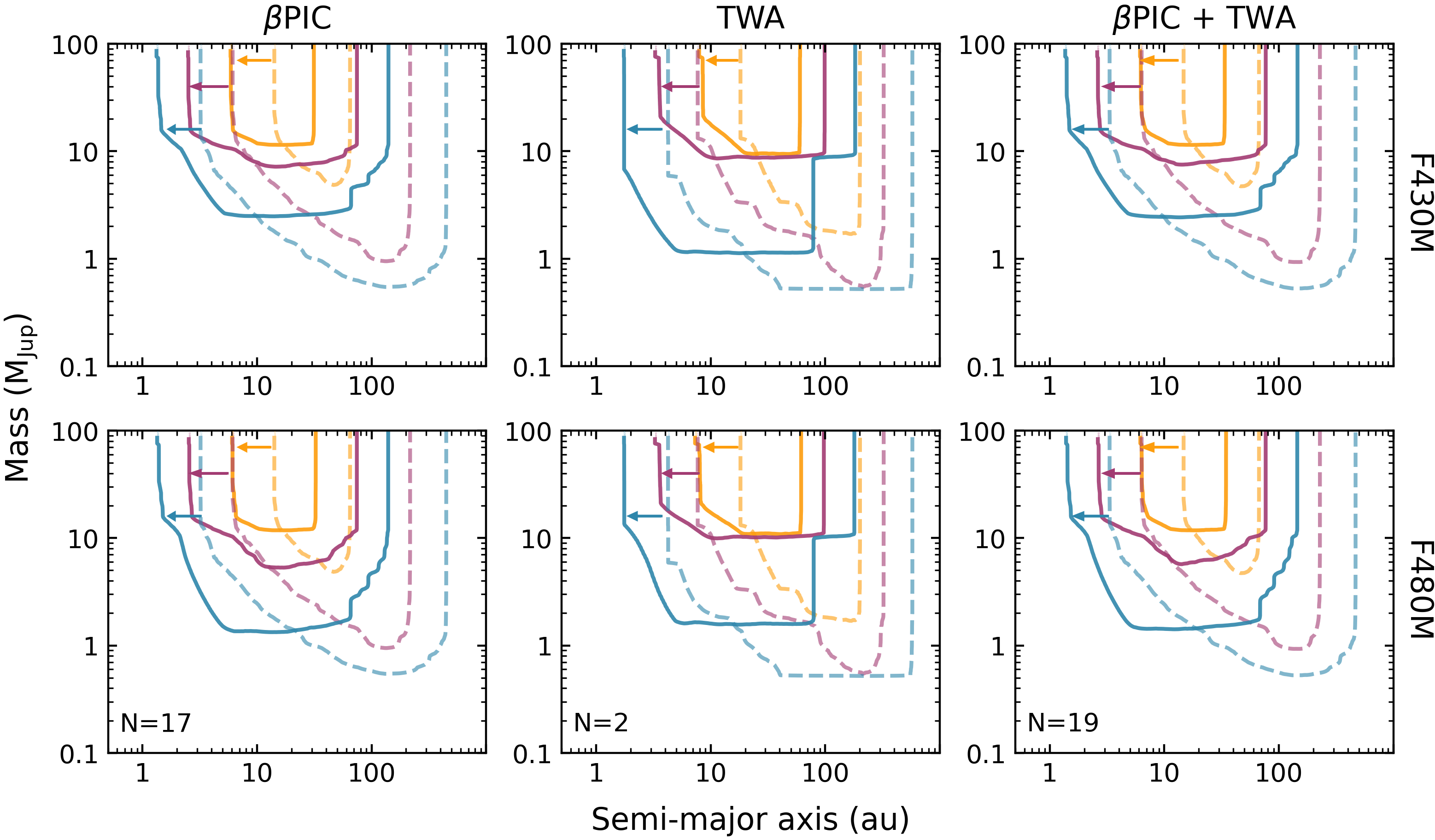}
    \caption{Direct comparison of the average detection probabilities of the common members from our sample and the SHINE survey for the $\beta$Pic and TWA moving groups. $N$ is the number of stars averaged over to obtain the probability in each case. The first and the second rows show the average probabilities of the common stars using the calculated F430M and F480M filters, respectively. The orange, pink and cyan colours show the 95\%, 68\% and 30\% contours respectively. The solid and dashed lines are for \textit{JWST}/\textit{NIRISS}/\textit{AMI} and SHINE respectively. The arrows show the shift in coverage of the innermost achievable separation between SPHERE and JWST.}
    \label{fig:SHINECrossmatch}
\end{figure*}

\section{Results and Discussion}
\label{sec:Discussion}

To understand the exquisite mass sensitivity limits attainable using the AMI mode with \textit{JWST}/\textit{NIRISS}, we present the detection probability maps of the total sample separated by spectral class (Figure \ref{fig:SpectralClassDetProb}), a cumulative average of the stars from each sample with maximum likelihood of a detection, separated by confidence levels (Figure \ref{fig:ConfidenceLevels}) and finally a direct comparison with ground-based instruments (Figure \ref{fig:SHINECrossmatch}).  

\subsection{Spectral Class}

Detection probabilities averaged over each of the spectral classes of the members in the sample provide insight into which type of stars are the most promising for detecting companions in a broader context. Figure \ref{fig:SpectralClassDetProb} separates the members of $\beta$Pic, TWA and TAA into the spectral groups of A, F/G/K and M stars. The samples of $\beta$Pic and TWA also have very few earlier type stars causing the probability contours to be slightly discontinuous. In Figure \ref{fig:SpectralClassDetProb} filled contours in the left, middle and right columns, show the average probabilities of the stars in each sample in the spectral groups A, F/G/K and M, in the F430M filter and the dotted lines 
show the same in the F480M filter. The value of $N$ shows the number of stars in each sample in the particular spectral classifications at different probabilities. The contours shown by the four dashed lines and the four subsequently darker regions in each plot are for confidences of 10\%, 50\%, 68\% and 95\%. We infer from this result that broadly, the F480M filter clearly outperforms the F430M filter when the aim is to access the lower mass companions. This arises from the fact that substeallar companions with lower masses (and thus cooler temperatures) have a greater fraction of their luminosity at longer wavelengths, making them brighter at $4.8\,\rm{\mu m}$ (central wavelength of F480M) than at $4.3\,\rm{\mu m}$ (central wavelength of the F430M filter). There is a clear pattern of increasing depth in each sample towards later spectral types. This is due to the later spectral types being dimmer and hence lower mass companions being potentially more accessible to detect due to a more favourable contrast. The other pattern that emerges from this result is the shift of contour lines outwards (i.e. further away from the host star), {with increasing median stellar group distance (see Figure \ref{fig:DistanceHist})}, for each spectral type from the samples of $\beta$Pic to TWA and then TAA.

{M stars dominate the stellar mass distribution in all the groups. This makes the average detection probability of the M stars in each group, a reasonable proxy for the group as a whole. For example, in Table \ref{table:yields}, for the case of the yield values with the \cite{Vigan2021A&A} distribution in the F480M filter, the mean values of all the stars, and only the M stars, have values 0.16 and 0.18 respectively in the case of TWA. These values are the closest to each other when compared to the mean values of A and F/G/K stars for the same distribution and filter for TWA, which are 0.08 and 0.04 respectively. This trend is seen for all groups, filters and distributions in Table \ref{table:yields}.} The 10\% confidence contour of the detection probabilities in the F480M filter, reaches masses of ${\sim}0.3{-}0.5\,\rm{M\textsubscript{Jup}}$, in the samples of $\beta$Pic, TWA and TAA, at separations ${\sim}3{-}5\,\rm{au}$. {Hence, preferentially selecting M stars to observe gives access to lower mass companions, and therefore more effectively taps into the distribution of planets as predicted by \cite{2021Fulton} and \cite{Vigan2021A&A}}. However, this result is for the average of many targets, and not necessarily an optimised list. In the next section, we show how selecting optimal targets can boost the detection efficiency.


\subsection{Best targets to detect close-in companions}
\label{ssec:CloseInSeparations}
Since the F480M filter provides better performance in terms of reaching lower mass limits, the results presented going forward to determine the best targets to detect close-in companions have been limited to this filter. In Figure \ref{fig:ConfidenceLevels}, we present these results. Ranked by the yield values of each of the members of the samples, obtained using the extended RV distribution (see section \ref{ssec:YeildCalculation}), the first, second and third rows show the averaged detection probabilities for the best 5, 15 and total number of targets respectively (values of $N$ in Figure \ref{fig:ConfidenceLevels}), for the members of each sample. The yield from the Extended RV distribution is used rather than the Bimodal distribution to rank the stars since the former dominates the close-in separation parameter space (see Figure \ref{fig:RankingPlot}) which is a better descriptor of the region of parameter space we are concerned with in this work (see section \ref{sec:Intro}), even though the latter produces higher yields. The left, centre and right columns show the 30\%, 68\% and 95\% confidence contours for each sample, including the SHINE survey, shown with a dashed contour. These yield values averaged over $N$ stars in Figure \ref{fig:ConfidenceLevels} are also presented for each sample in the right column. 

The H\textsubscript{2}O and CO frost lines for the stars in our sample were calculated using a methodology from \cite{Vigan2021A&A}, which calculates the extent of the frost line using the evaporation temperatures ($135\,\rm{K}$ and $20\,\rm{K}$  for H\textsubscript{2}O and CO respectively) from \cite{2011Oberg}, a parametric disk temperature profile from \cite{1974Lewis} and observations of protoplanetary disks from \cite{2005Andrews,2007Andrews2,2007Andrews}. Since frost lines have uncertainties on them, a gradient region demarcated by the smallest and the largest separation values from the range of calculated values for our stars is plotted in Figure \ref{fig:ConfidenceLevels}. The darkest region is the halfway point between the two extremities and the gradient decreases linearly on either side. This is represented in a logarithmic scale in the figure.


{\textit{For the goal of detecting sub-Jupiter mass companions near the water frost lines at the  68\% confidence level, only the most favourable five stars (or 15 stars to some extent) in the $\beta$Pic and TWA moving groups should be targeted}, as evident from Figure \ref{fig:ConfidenceLevels}.} In addition to this, the companions with masses greater than {${\sim}1\,\rm{M_{Jup}}$} near and exterior to these separations can be detected around the best 5 and 15 stars for all the groups {(including TAA)} with a confidence of 68\%. This is particularly remarkable since at higher masses (${\sim}5{-}10\,\rm{M_{Jup}}$), the variation of luminosities in the hot and cold start models is more pronounced \citep{2012Spiegel,Wallace2021}. 

{The very low infrared background offered by \textit{JWST} allows impressive sensitivity to low mass companions (e.g. $1{-}2\,\rm{M_{Jup}}$), even at 95\% confidence in some cases (right column in Figure \ref{fig:ConfidenceLevels}), as well as for a majority of the stars in each stellar group (bottom row in Figure \ref{fig:ConfidenceLevels}).}

The \textit{Gaia} mission is expected to unveil thousands of planets \citep[][]{2014Sozzetti} with reasonably well constrained masses. Most of these will be at separations within ${\sim}10\,\rm{au}$, and as we have demonstrated, \textit{JWST/AMI} can image companions at these locations and measure their mid-infrared luminosities. 
{A tightly constrained dynamical mass, combined with the precise estimate of the bolometric luminosity that can be delivered with \textit{JWST/NIRISS/AMI}, can then place powerful constraints on the initial energy budget of the companion, and the degree to which it has been modifed due to e.g., energy losses due to accretion shocks at the surface of the planet \citep[e.g.,][]{mfh07,Marleau2014}.}
For example, as in Table 1 in \cite{2012Spiegel}, a $10\,\rm{Myr}$ old planet of ${\sim}1\,\rm{M_{Jup}}$, would have ${\sim}2$ times higher luminosity in the hot start model versus a cold start scenario. {A $10\,\rm{M_{Jup}}$ planet at the same age would have a hot-start luminosity ${\sim}35$ times that of a cold-start luminosity. Hence, the population of planets to which AMI is sensitive would be an excellent indicator of initial entropies. As an example, to better understand this approach, the top panel of Figure \ref{fig:EntropyCalculation} shows hypothetical detections of $2\,\rm{M\textsubscript{jup}}$ and $10\,\rm{M\textsubscript{jup}}$ mass planets in $\beta$Pic and TAA, at $3\,\rm{au}$ and $100\,\rm{au}$ respectively, given the group specific detection probability maps. The bottom panel of Figure \ref{fig:EntropyCalculation} shows how given the different ages of  $\beta$Pic and TAA, different luminosities (in M band) would hint at different initial entropies \citep[recreated from][]{2012Spiegel}. As can be seen in the figure, this difference in initial entropy is more pronounced if the companions are younger or if they are more massive.}




The cumulative average yield values in Figure \ref{fig:ConfidenceLevels}, decrease as the value of $N$ increases since we are averaging over stars which have lower probabilities of hosting companions in the region of the RV distribution. The 95\% contour for the $N{=}N_{tot}$ case provides the expected shape of the relative detection probabilities for the members of $\beta$Pic, TWA and TAA because of the subsequent decrease in age (hence the contours go consequently deeper) and the increase in average distance (hence they are restricted to wider orbital separations). This trend is absent in the $N{=}5$ and $N{=}15$ cases since the stars hence selected are the ones with the highest yields from their parent samples and have a broad range of distances and masses (see Figures \ref{fig:DistanceHist} and \ref{fig:MassHist} respectively). This trend also does not manifest in the 68\% and the 30\% confidence contours since at these confidences, the limiting factor is primarily the sensitivity of the instrument itself, rather than the properties of the stellar groups. {Figure \ref{fig:ConfidenceLevels} also gives a coarse comparison of the performance of the AMI mode with \textit{JWST}/\textit{NIRISS} compared to the results from the SHINE survey on the dedicated ground based \textit{VLT}/\textit{SPHERE} instrument \citep{Vigan2021A&A}}.  However, Figure \ref{fig:ConfidenceLevels} does not present a fully fair comparison since the same set of stars between the two surveys are not being compared. Rather, this exercise compares the best targets from our sample with the entirety of the SHINE survey. So, to give a fairer comparison, we present the results of a direct comparison with the SHINE survey in the following section.

\subsection{Direct comparison with the \textit{SHINE} survey}

{Figure \ref{fig:SHINECrossmatch} shows the averaged probability maps for those stars in $\beta$Pic (17 such stars) and TWA (2 such stars) that are common to our sample as well as the SHINE sample.} The number of stars considered is given by the $N$ value in the plots. None of the stars in TAA were observed with SHINE, most likely since these have declinations too far North to be observed from the southern location of \textit{VLT}. In the Figure, the 95\%, 68\% and 30\% confidences of the mean detection probabilities are averaged over $N$ stars. The solid and the dashed lines represent the \textit{JWST}/\textit{NIRISS} and the SHINE survey contours respectively. The first and the second rows show the results from the F430M and F480M filters respectively. {The first and the second column show the $\beta$Pic and TWA stars respectively which are common to both samples}.

The right most column shows the average of all 19 cross-matched stars. It is evident from this result that SHINE reaches slightly deeper contrasts when compared to the AMI mode (with the F430M and F480M filters) at larger separations. However, the technical edge achieved by the latter is the accessibility of the regions closer to the host star. In the second row in Figure \ref{fig:SHINECrossmatch} (F480M filter), the inner limit of the 95\% contour for the averaged detection probability for $\beta$Pic members is brought down from over ${\sim}10\,\rm{au}$ (with SHINE) to only ${\sim}5\,\rm{au}$ (with \textit{JWST}/\textit{NIRISS}/\textit{AMI}) for companions with masses ${>}10\,\rm{M\textsubscript{Jup}}$. Similar improvements are seen when looking at the TWA averaged members as well as the average of all members from $\beta$Pic and TWA, across both the filters. This spatial improvement is marked by colour coded arrows in the plots. This makes \textit{JWST} the ideal observatory to perform a survey for substellar objects near the {circumstellar frost lines of nearby stars}, since it can achieve a combination of sensitivity at mid-infrared wavelengths and accessibility to close-in separations with better inner working angles in the AMI mode.

Our demonstration that \textit{NIRISS} operating in AMI mode achieves superior sensitivity at closer orbital separations than \textit{SPHERE} \textit{for the same set of stars} is particularly noteworthy given that \textit{JWST} utilizes a smaller telescope aperture than the one used by \textit{VLT} ($6.5\,\rm{m}$ versus $8\,\rm{m}$), as well as operating at a longer wavelength (${\sim}4.8\,\rm{\mu m}$ for \textit{JWST}/\textit{NIRISS} versus 1-2$\,\rm{\mu m}$ for \textit{VLT}/\textit{SPHERE}), an observational configuration that would indeed return a poorer {inner working angle} in the case of conventional coronagraphic imaging.  This superior performance relative to \textit{VLT}/\textit{SPHERE} is due to the interferometric configuration utilized in the AMI mode. In addition to this, observations in the ${\sim}3{-}5\,\rm{\mu m}$ region of the spectrum is extremely important for complementing measurements from observations made by the instruments \textit{GPI}, \textit{VLT/SPHERE} and \textit{VLTI/GRAVITY} at ${\sim}1{-}2\,\rm{\mu m}$. {The long wavelength coverage can provide a much better estimate of the overall bolometric luminosity of the object, which is likely a more secure value from which to draw conclusions about intial entropies. The ${\sim}3{-}5\,\rm{\mu m}$ wavelength range is also particularly well suited to discriminate atmospheric models that incorporate various levels of disequilibrium chemistry that could be due to dynamical processes such as vertical atmospheric mixing  \citep{2012Skemer,PhillipsPaper}. Differentiating changes in the spectrum that could be induced by such dynamical atmospheric processes from those caused by differences in chemical abundances will ultimately allow tighter constraints to be placed on the intrinsic chemical composition of planetary atmospheres \citep[e.g.][]{cbs07}} 



\subsection{Kernel phase performance with \textit{JWST}}
\label{ssec:kernelPhase}
\begin{figure}
\centering
	\includegraphics[width=\columnwidth]{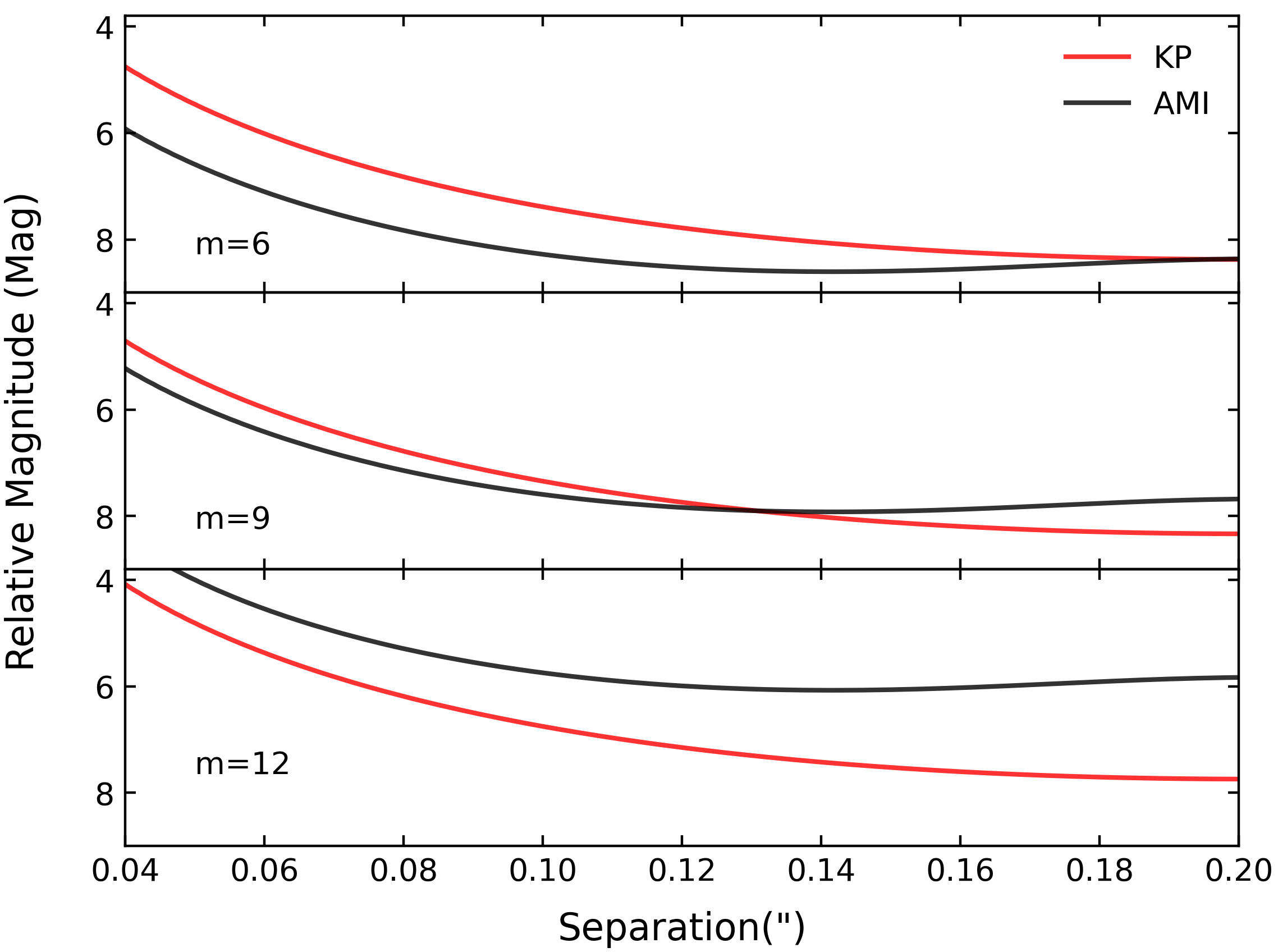}
    \caption{{Contrast curves for aperture masking interferometry (AMI, in red) and kernel phase (KP, in black) cases for apparent magnitude values ($m$) of 6, 9 and 12 respectively, at $4.8\,\rm{\mu m}$. For stars with m$\gtrsim$9, KP clearly outperforms AMI in terms of reaching deeper contrasts for a fixed integration time. So, AMI is recommended to be used for stars with m$\lesssim$9 to detect companions around them.}}
    \label{fig:KPvsAMI}
\end{figure}

{In the high-Strehl regime, the interferometric technique of KP \citep[][]{2010Martinache} represents a viable alternative to AMI. Both AMI and KP observations in \cite{StephJatis} were simulated with a fixed exposure time of six hours for each observation (see section \ref{ssec:ContrastCurves}). For fainter targets, AMI requires more integration time compared to KP to reach equivalent contrasts. Hence in this scenario of observing fainter stars, KP outperforms AMI in terms of achieving higher sensitivities for targets with apparent magnitudes ${m{\gtrsim}9}$. This is shown in Figure \ref{fig:KPvsAMI}. For apparent magnitudes of $m$ = 6, 9, and 12, the figure shows the contrast curves for simulated observations using KP and AMI with a fixed integration time of six hours. AMI clearly reaches deeper contrasts for brighter targets ($m{\lesssim}9$, for example $m{=}6$ in the figure) and KP reaches deeper contrasts for fainter targets ($m{\gtrsim9}
$, for example $m{=}12$ in the figure). However, while planning an actual survey, for bright targets, AMI would not necessarily require six hours for each target and visits can be optimised on a case-by-case basis to achieve similar contrasts and mass ranges presented in sections \ref{ssec:ContrastCurves} and \ref{ssec:CloseInSeparations} respectively, with lesser exposure times. An actual survey would potentially use both KP and AMI observations for improved efficiency, depending on the brightness of the targets.}

\subsection{Distinguishing between planetary populations}
{Using the yield values in Table \ref{table:yields} in conjunction with future survey with \textit{JWST/NIRISS/AMI}, attempts can be made towards distinguishing planetary populations. For example, if observing 20 stars in the TWA sample results in $\sim$three companions detections, the bimodal population is more likely to be prevalent in this moving group ($20{\times}0.16{\approx}3$, where 0.16 is the mean yield value with the bimodal population described in \citealt{Vigan2021A&A}). On the contrary if observing 20 stars in the TWA sample results in $\sim$one companion detection, the underlying population would most likely be better described by the RV population ($20{\times}0.07{\approx}1$, where 0.07 is the mean yield value with the \citealt{2021Fulton}/RV population).}

\section{Conclusions}
\label{sec:Conclusions}
We have presented in this work the exquisite capabilities of the AMI mode using  \textit{JWST}/\textit{NIRISS} to image Jupiter and sub-Jupiter mass exoplanets near the water frost-lines around nearby young stars. {Both $\beta$ Pictoris and TW Hydrae moving groups host ${\sim}10\,\rm{stars}$ each to image sub-Jupiter companions with very high confidences (${\sim}68\%$).} {This is a consequence of the  \textit{JWST/AMI} mode being able to achieve contrasts of ${\sim}10^{-4}$ at separations of $\lambda/D$ and wider, with sufficient integration times \citep[][]{2020SoulainSPIE}}. A future survey with this mode to target these detectable planets to put constraints on early entropy conditions of planet formation can be executed in conjunction with a coronagraphy survey of the same stars to save telescope time. 

{Picking the 10-15 best targets (as shown in Figure \ref{fig:ConfidenceLevels}) either from TWA or $\beta$Pic, a survey of such stars would take a total of $60{-}90\,\rm{hours}$ assuming a fixed exposure time of $6\,\rm{hours}$ on each target (see \S\ref{ssec:ContrastCurves}). However, targets which are brighter, would not require as much time to gain the required SNR for a detection and hence this estimated survey time is only an upper limit.}

In addition to this, the mode also achieves very high yields for detecting companions in general across the stellar groups, which points to the lucrative nature of a future \textit{JWST} exoplanet survey with AMI. {For example, even the least mean yield values in Table \ref{table:yields} of 0.04-0.05, is greater than most ground based surveys to date, which have values converging at ${\sim}0.01$ \citep[][]{2018Bowler}. And the highest mean yield value in Table \ref{table:yields} is 0.16 which is ${\sim}4$ times the minimum value. An optimised survey picking the best candidate stars would have yield values even greater still (see yield values in Tables \ref{sec:BPICStars}, \ref{sec:TWAStars} and \ref{sec:TAAStars} in the appendix).} 


Limiting such a survey to the stars of the Taurus-Auriga association would significantly reduce observatory overheads compared to a survey of $\beta$ Pictoris and TW Hydrae, due to the members of the former being close to each other on the sky plane. Using the stars in a sequence of observations such that they work as as set of mutual reference stars would go a step further in constraining the elapsed time of such a survey. The yield calculations indicate $\sim$0.05 detections per star for the association, {which is more than the yield of most ground based surveys carried out to date} \citep{2019Nielsen,Vigan2021A&A}. Ground based high contrast imaging platforms with visible-light wave front sensors will not typically be effective for these targets due to their faint optical magnitudes, making \textit{JWST} the ideal observatory for this task. However, the efficiency of a survey in TAA could be impacted by other variables such as: (1) the presence of protoplanetary discs, which could potentially obscure forming planets; or (2) not carrying out the observations in a non-interruptible sequence which would result in increased telescope overheads. {Non-interruptible observations is a mode offered by \textit{JWST}, which enables the observer to carry out a sequence of observations in a specified time. This will not only save on telescope slew time by optimizing the sequence in order of the closest stars on the sky for the telescope to point, but will also ensure any drift in wave front error, for example due to thermal/structural evolution of the telescope will be minimized.}

The upcoming sequential data releases from the \textit{Gaia} mission are expected to unveil hundreds, if not thousands, of planets orbiting nearby stars in the vicinity of the frost lines of these stars \citep[e.g.,][]{2014Sozzetti}. Some fraction of these stars will have ages ${\lesssim}100\,\rm{Myr}$, and thus will potentially host companions sufficiently self-luminous to be suitable for direct imaging. However, even orbital separations of $2{-}3\,\rm{au}$ for a star at $50\,\rm{pc}$ correspond to angular separations of ${\sim}40{-}50\,\rm{mas}$, which is comparable to the resolution limit of $8{-}10\,\rm{m}$ telescopes operating in the near-infrared. {In this paper we have also demonstrated that \textit{JWST} operating in the AMI mode has comparable sensitivities and inner working angles as \textit{VLTI} instruments at $1{-}2\,\rm{\mu m}$ {\citep[][]{2019Lacour,2020Nowak,2022HinkleyA}}, but \textit{crucially} provides complementary wavelength coverage at $3{-}5\,\rm{\mu m}$, which is an advantageous wavelength region for discriminating SED shapes that are driven by changes in intrinsic composition, and SED shapes that are being affected by atmospheric processes that lead to disequilibrium chemistry, like vertical atmospheric mixing \citep{2012Skemer,2013Konopacky,PhillipsPaper,2020Miles}.}

Lastly, we await the release of the first science observations from \textit{JWST}, which would enable us to better understand the contrast limits with the AMI mode, compared to the simulations. This is one of the goals of Director’s Discretionary Early Release Science Program 1386, \textit{High Contrast Imaging of Exoplanets and Exoplanetary Systems with JWST} {\citep[][]{2022HinkleyB}}, with which the AMI observation would serve as the benchmark for future observations in this mode and {evaluate on-sky} contrasts and hence detection probabilities.

\section*{Acknowledgements}

 We thank Adam Kraus for valuable discussion on the stars in the TAA sample. We thank Arthur Vigan for providing us with the methodology to calculate the positions of the frost lines. {We thank Ken Rice for useful discussions of planet formation models.} We also thank the anonymous referee whose comments have been invaluable towards improving this paper. SR is supported by a Global Excellence Award by the University of Exeter. ALC is supported by a grant from STScI (\textit{JWST}-ERS-01386) under NASA contract NAS5-03127.

\section*{Data Availability}

The data underlying this article will be shared on reasonable request
to the corresponding author.



\bibliographystyle{mnras}
\bibliography{example, MasterBiblio_Sasha.bib} 




\appendix

\section{Contrast curve simulation variables}

\begin{table*}

\begin{tabular}{c | c c c c c | c c c c c} 
\hline\hline 

&&& \textbf{F430M} &&&&& {F480M} \\
\hline

M\textsubscript{s} & $n_{g}$ & $n_{int}$ & $n_{g,rem}$ & $t_{tot} (s)$ & \textit{eff} & $n_{g}$ & $n_{int}$ & $n_{g,rem}$ & $t_{tot} (s)$ & \textit{eff}\\
\hline 
5.7 & 35 & 2070 & 31 & 5246 & 0.97 & 39 & 1858 & 19 & 5261 & 0.97\\
5.8 & 38 & 1907 & 15 & 5258 & 0.97 & 42 & 1725 & 31 & 5271 & 0.98\\
5.9 & 42 & 1725 & 31 & 5271 & 0.98 & 46 & 1575 & 31 & 5282 & 0.98\\
6.0 & 46 & 1575 & 31 & 5282 & 0.98 & 51 & 1421 & 10 & 5294 & 0.98\\
6.1 & 50 & 1449 & 31 & 5292 & 0.98 & 56 & 1294 & 17 & 5303 & 0.98\\
6.2 & 55 & 1317 & 46 & 5302 & 0.98 & 61 & 1188 & 13 & 5311 & 0.98\\
6.3 & 61 & 1188 & 13 & 5311 & 0.98 & 67 & 1081 & 54 & 5319 & 0.99\\
6.4 & 67 & 1081 & 54 & 5319 & 0.99 & 74 & 979 & 35 & 5327 & 0.99\\
6.5 & 73 & 992 & 65 & 5326 & 0.99 & 81 & 894 & 67 & 5333 & 0.99\\
6.6 & 80 & 906 & — & 5332 & 0.99 & 89 & 814 & 35 & 5339 & 0.99\\
6.7 & 88 & 823 & 57 & 5339 & 0.99 & 98 & 739 & 59 & 5345 & 0.99\\
6.8 & 96 & 755 & — & 5344 & 0.99 & 107 & 677 & 42 & 5349 & 0.99\\
6.9 & 106 & 683 & 83 & 5349 & 0.99 & 117 & 619 & 58 & 5354 & 0.99\\
7.0 & 116 & 624 & 97 & 5353 & 0.99 & 129 & 561 & 112 & 5358 & 0.99\\
7.1 & 127 & 570 & 91 & 5357 & 0.99 & 141 & 514 & 7 & 5362 & 0.99\\
7.2 & 140 & 517 & 101 & 5361 & 0.99 & 155 & 467 & 96 & 5365 & 0.99\\
7.3 & 153 & 473 & 112 & 5365 & 0.99 & 170 & 426 & 61 & 5368 & 0.99\\
7.4 & 168 & 431 & 73 & 5368 & 0.99 & 186 & 389 & 127 & 5371 & 0.99\\
7.5 & 184 & 393 & 169 & 5371 & 0.99 & 205 & 353 & 116 & 5374 & 1.00\\
7.6 & 202 & 358 & 165 & 5373 & 1.00 & 224 & 323 & 129 & 5376 & 1.00\\
7.7 & 221 & 327 & 214 & 5375 & 1.00 & 246 & 294 & 157 & 5378 & 1.00\\
7.8 & 243 & 298 & 67 & 5378 & 1.00 & 270 & 268 & 121 & 5380 & 1.00\\
7.9 & 266 & 272 & 129 & 5380 & 1.00 & 296 & 244 & 257 & 5382 & 1.00\\
8.0 & 292 & 248 & 65 & 5381 & 1.00 & 324 & 223 & 229 & 5383 & 1.00\\
8.1 & 320 & 226 & 161 & 5383 & 1.00 & 356 & 203 & 213 & 5385 & 1.00\\
8.2 & 351 & 206 & 175 & 5384 & 1.00 & 390 & 185 & 331 & 5386 & 1.00\\
8.3 & 385 & 188 & 101 & 5386 & 1.00 & 428 & 169 & 149 & 5387 & 1.00\\
8.4 & 422 & 171 & 319 & 5387 & 1.00 & 469 & 154 & 255 & 5388 & 1.00\\
8.5 & 463 & 156 & 253 & 5388 & 1.00 & 514 & 141 & 7 & 5389 & 1.00\\
8.6 & 508 & 142 & 345 & 5389 & 1.00 & 564 & 128 & 289 & 5390 & 1.00\\
8.7 & 557 & 130 & 71 & 5390 & 1.00 & 618 & 117 & 175 & 5391 & 1.00\\
8.8 & 611 & 118 & 383 & 5391 & 1.00 & 678 & 106 & 613 & 5392 & 1.00\\
8.9 & 670 & 108 & 121 & 5392 & 1.00 & 743 & 97 & 410 & 5393 & 1.00\\
9.0 & 735 & 98 & 451 & 5393 & 1.00 & 800 & 90 & 481 & 5393 & 1.00\\
9.1-12.7& 799 & 90 & 571 & 5393 & 1.00 & 800 & 90 & 481 & 5393 & 1.00\\

\hline 
\end{tabular}
\caption{\textit{JWST/NIRISS/AMI} simulated observation details recreated from \protect\cite{StephJatis}, in the filters F430M and F480M. M\textsubscript{s} is the apparent magnitude of the target star and is listed in increments of 0.1; $n_{g}$ is the number of groups in each integration; $n_{int}$ is the number of integrations in 90 minutes; $n_{g,rem}$ is the additional number of groups (in one final shorter integration to finish a 90 minute observing block);  $t_{tot}$ is the total integration time in 90 minutes; \textit{eff} is the observing efficiency rounded to 0.01.} 
\label{table:Simulations} 
\end{table*}

\section{List of stars in the sample}

\begin{table*}
\label{BPICappendix}
	\centering
	\begin{tabular}{rccccccc} 
		\hline
		\hline
		Rank & \textit{Gaia} DR2 ID & Distance (pc) & Spectral Type & $m\textsubscript{F430M}$ & $m\textsubscript{F480M}$ & $y\textsubscript{RV}$ & $\overline{y}\textsubscript{RV}$\\
		\hline
        1 & 3230008650057256960 & 21.00 & M9 & 11.93 & 12.16 & 0.147 & 0.147\\
        2 & 5355751581627180288 & 19.79 & M5 & 8.64 & 8.82 & 0.129 & 0.138\\
        3 & 3238965099979863296 & 27.63 & M4 & 9.08 & 9.24 & 0.115 & 0.130\\
        4 & 6603693881832177792 & 20.87 & M4 & 7.87 & 8.03 & 0.115 & 0.127\\
        5 & 2901786974419551488 & 29.76 & M4 & 9.27 & 9.43 & 0.113 & 0.124\\
        6 & 6794047652729201024 & 9.71 & M1 & 5.23 & 5.37 & 0.112 & 0.122\\
        7 & 2324205785406060928 & 37.38 & M6 & 11.65 & 11.84 & 0.109 & 0.120\\
        8 & 3216753556349327232 & 38.52 & M5 & 10.67 & 10.84 & 0.102 & 0.118\\
        9 & 3291643148740384128 & 23.79 & M2 & 7.26 & 7.42 & 0.100 & 0.116\\
        10 & 3216729878197029120 & 38.02 & M5 & 9.51 & 9.68 & 0.100 & 0.114\\
        11 & 2727844441062478464 & 37.39 & M5 & 10.01 & 10.18 & 0.100 & 0.113\\
        12 & 2315841869173294080 & 35.03 & M3 & 8.97 & 9.12 & 0.098 & 0.112\\
        13 & 2433191886212246784 & 27.45 & M0 & 7.57 & 7.71 & 0.098 & 0.111\\
        14 & 3231945508509506176 & 24.40 & M0 & 7.23 & 7.36 & 0.097 & 0.110\\
        15 & 2477870708709917568 & 37.28 & M4 & 9.12 & 9.28 & 0.097 & 0.109\\
        16 & 6833291426043854976 & 33.60 & M5 & 8.50 & 8.66 & 0.096 & 0.108\\
        17 & 6577998398172195840 & 48.72 & M5 & 11.75 & 11.94 & 0.096 & 0.107\\
        18 & 4707563810327288192 & 36.82 & M3 & 8.77 & 8.92 & 0.095 & 0.107\\
        19 & 4764027962957023104 & 26.87 & M0 & 7.22 & 7.35 & 0.094 & 0.106\\
        20 & 2899492637251200512 & 33.77 & M3 & 8.14 & 8.30 & 0.093 & 0.105\\
        21 & 6800238044930953600 & 43.66 & M4 & 9.80 & 9.97 & 0.093 & 0.105\\
        22 & 68012529415816832 & 50.70 & M8 & 12.28 & 12.56 & 0.091 & 0.104\\
        23 & 3216729573251961856 & 36.74 & M1 & 8.02 & 8.16 & 0.089 & 0.104\\
        24 & 6806301370519190912 & 43.88 & M4 & 8.76 & 8.92 & 0.085 & 0.103\\
        25 & 6649786646225001984 & 51.65 & M4 & 10.59 & 10.72 & 0.083 & 0.102\\
        26 & 6382640367603744128 & 36.72 & K7 & 7.91 & 8.04 & 0.083 & 0.101\\
        27 & 132362959259196032 & 40.94 & K7 & 8.12 & 8.23 & 0.081 & 0.101\\
        28 & 5266270443442455040 & 39.11 & K4 & 7.72 & 7.82 & 0.077 & 0.100\\
        29 & 5935776714456619008 & 50.79 & M3 & 8.69 & 8.82 & 0.071 & 0.099\\
        30 & 6736232346363422336 & 49.46 & K8 & 8.55 & 8.68 & 0.070 & 0.098\\
        31 & 6747467224874108288 & 51.31 & K9 & 8.91 & 9.04 & 0.068 & 0.097\\
        32 & 3393207610483520896 & 53.09 & K2 & 8.69 & 8.81 & 0.067 & 0.096\\
        33 & 87555176071871744 & 70.75 & M6 & 12.56 & 12.75 & 0.066 & 0.095\\
        34 & 4067828843907821824 & 63.81 & M2 & 9.52 & 9.67 & 0.065 & 0.094\\
        35 & 2622845684814477696 & 25.52 & F8 & 5.21 & 5.25 & 0.065 & 0.093\\
        36 & 3009908378049913216 & 26.84 & F8 & 5.45 & 5.49 & 0.065 & 0.092\\
        37 & 5882581895219921024 & 38.72 & K0 & 6.29 & 6.36 & 0.059 & 0.092\\
        38 & 94988050769772288 & 52.77 & K0 & 7.62 & 7.73 & 0.058 & 0.091\\
        39 & 5811866422581688320 & 30.35 & K1 & 5.27 & 5.34 & 0.058 & 0.090\\
        40 & 6655168686921108864 & 47.25 & G9 & 7.08 & 7.14 & 0.058 & 0.089\\
        41 & 5945104588806333824 & 76.64 & M2 & 10.29 & 10.44 & 0.057 & 0.088\\
        42 & 6663346029775435264 & 71.27 & M0 & 9.31 & 9.44 & 0.054 & 0.087\\
        43 & 6643589352010758400 & 47.78 & F6 & 6.40 & 6.42 & 0.052 & 0.087\\
        44 & 5924485966955008896 & 67.61 & K1 & 8.25 & 8.32 & 0.051 & 0.086\\
        45 & 6847146784384459648 & 50.11 & F5 & 6.44 & 6.47 & 0.051 & 0.085\\
        46 & 6882840883190250752 & 45.91 & F8 & 6.31 & 6.34 & 0.050 & 0.084\\
        47 & 3205095125321700480 & 29.91 & F0 & 4.83 & 4.85 & 0.049 & 0.084\\
        48 & 6760846563417053056 & 74.34 & M0 & 8.99 & 9.12 & 0.049 & 0.083\\
        49 & 4792774797545105664 & 19.63 & A6 & 3.70 & 3.72 & 0.049 & 0.082\\
        50 & 4045698423617983488 & 71.48 & K5 & 8.17 & 8.28 & 0.047 & 0.081\\
        51 & 6631762764424312960 & 50.57 & F5 & 6.62 & 6.65 & 0.046 & 0.081\\
        52 & 6438274350302427776 & 28.79 & A7 & 4.58 & 4.61 & 0.046 & 0.080\\
        53 & 107774202769886848 & 39.56 & F5 & 5.50 & 5.53 & 0.046 & 0.079\\
        54 & 5946515438335508864 & 65.80 & F8 & 7.63 & 7.67 & 0.045 & 0.079\\
        55 & 6470519830886970880 & 63.67 & F5 & 7.06 & 7.09 & 0.045 & 0.078\\
        56 & 6702775135228913280 & 49.30 & F6 & 6.24 & 6.27 & 0.044 & 0.078\\
        57 & 5849837854817580672 & 16.40 & A7 & 2.47 & 2.50 & 0.042 & 0.077\\
        58 & 4051081838710783232 & 80.48 & G5 & 7.76 & 7.82 & 0.036 & 0.076\\
        59 & 6724105656508792576 & 43.97 & A6 & 4.65 & 4.68 & 0.035 & 0.075\\
        60 & 4038504701367019648 & 82.71 & G0 & 7.77 & 7.82 & 0.034 & 0.075\\
        61 & 4057573802035360896 & 83.29 & F3 & 7.29 & 7.31 & 0.030 & 0.074\\

		\hline
	\end{tabular}
		\caption{Properties of the stars in the $\beta$Pic sample ranked by the RV yield value: The distances are in parsec. $m\textsubscript{F430M}$ and $m\textsubscript{F480M}$ are the apparent magnitudes in the F430M and the F480M filters respectively and are calculated using a method explained in section \ref{ssec:FilterMagnitudeCalculation}. $y\textsubscript{RV}$ is the yield for each star based off the radial velocity distribution when simulated to be observed with the F480M filter and is calculated as explained in section \ref{ssec:YeildCalculation}. $\overline{y}\textsubscript{RV}$ is the moving cumulative average of the $y\textsubscript{RV}$ values. {Please note that this list is produced from an interpolated parameter space of simulated contrast curves and so, while planning actual observations, individual targets should be simulated separately for improved accuracy.}}
		\label{sec:BPICStars}

\end{table*}

\begin{table*}
	\centering
	\begin{tabular}{rccccccc} 
		\hline
		\hline
		Rank & \textit{Gaia} DR2 ID & Distance (pc) & Spectral Type & $m\textsubscript{F430M}$ & $m\textsubscript{F480M}$ & $y\textsubscript{RV}$ & $\overline{y}\textsubscript{RV}$\\
		\hline
        1 & 3478519134297202560 & 46.71 & M8 & 12.24 & 12.50 & 0.106 & 0.106\\
        2 & 3536988276442796800 & 43.80 & M6 & 10.34 & 10.52 & 0.104 & 0.105\\
        3 & 3478940625208241920 & 48.90 & M5 & 9.81 & 9.99 & 0.101 & 0.104\\
        4 & 3481965141177021568 & 47.42 & M5 & 9.80 & 9.97 & 0.098 & 0.102\\
        5 & 6146137782994601984 & 52.93 & M3 & 10.19 & 10.34 & 0.093 & 0.100\\
        6 & 5396978667759696000 & 37.05 & M4 & 7.52 & 7.69 & 0.093 & 0.099\\
        7 & 3485098646237003392 & 45.96 & M3 & 8.51 & 8.66 & 0.088 & 0.097\\
        8 & 5460240959047928832 & 52.54 & M3 & 8.83 & 9.00 & 0.086 & 0.096\\
        9 & 5444751795151480320 & 34.10 & M2 & 7.93 & 10.52 & 0.084 & 0.095\\
        10 & 6150861598480158336 & 53.60 & M0 & 8.94 & 8.55 & 0.083 & 0.093\\
        11 & 5467714064704570112 & 61.17 & M5 & 10.22 & 10.36 & 0.082 & 0.092\\
        12 & 6146107993101452160 & 57.48 & M2 & 9.27 & 9.42 & 0.082 & 0.092\\
        13 & 5452498537466667776 & 45.94 & M2 & 7.74 & 7.88 & 0.081 & 0.091\\
        14 & 3465989374664029184 & 62.58 & M4 & 10.81 & 10.98 & 0.079 & 0.090\\
        15 & 3468438639892079360 & 64.35 & M5 & 10.58 & 10.76 & 0.078 & 0.089\\
        16 & 6147044433411060224 & 63.60 & M2 & 9.57 & 9.73 & 0.076 & 0.088\\
        17 & 5398663566249861120 & 49.67 & M2 & 7.72 & 7.87 & 0.075 & 0.087\\
        18 & 3466308095597260032 & 56.82 & M2 & 8.79 & 8.94 & 0.074 & 0.087\\
        19 & 5399220743767211776 & 59.85 & M1 & 8.71 & 8.84 & 0.071 & 0.086\\
        20 & 5396105586807802880 & 65.40 & M1 & 9.20 & 9.35 & 0.071 & 0.085\\
        21 & 5378040370245563008 & 72.25 & M0 & 10.06 & 10.22 & 0.069 & 0.084\\
        22 & 5401795662560500352 & 60.14 & K6 & 8.00 & 8.12 & 0.068 & 0.084\\
        23 & 3465944500845668224 & 70.76 & M4 & 9.67 & 9.84 & 0.068 & 0.083\\
        24 & 6132146982868270976 & 80.21 & M3 & 9.49 & 9.65 & 0.060 & 0.082\\
        25 & 3463395519357786752 & 76.49 & K5 & 8.78 & 8.90 & 0.057 & 0.081\\
        26 & 3532027383058513664 & 54.60 & A1 & 5.56 & 5.59 & 0.043 & 0.080\\
        27 & 6147117727029871360 & 70.77 & A0 & 5.92 & 5.94 & 0.032 & 0.078\\

		\hline
	\end{tabular}
		\caption{Same as in Table \ref{sec:BPICStars} but for the TWA sample}
		\label{sec:TWAStars}

\end{table*}

\begin{table*}
	\centering
	\begin{tabular}{rccccccc} 
		\hline
		\hline
		Rank & \textit{Gaia} DR2 ID & Distance (pc) & Spectral Type & $m\textsubscript{F430M}$ & $m\textsubscript{F480M}$ & $y\textsubscript{RV}$ & $\overline{y}\textsubscript{RV}$\\
		\hline
        1 & 3401526068784149504 & 58.34 & M0 & 9.47 & 9.64 & 0.101 & 0.101\\
        2 & 3403016495451584000 & 102.08 & K4 & 9.51 & 9.65 & 0.055 & 0.078\\
        3 & 146764465639042176 & 125.48 & M7 & 11.96 & 12.13 & 0.048 & 0.068\\
        4 & 146277553787186048 & 126.77 & M7 & 12.16 & 12.34 & 0.046 & 0.062\\
        5 & 3416236744087968768 & 118.44 & K7 & 9.77 & 9.92 & 0.046 & 0.059\\
        6 & 151028990206478080 & 126.97 & M6 & 11.50 & 11.67 & 0.046 & 0.057\\
        7 & 147799209159857280 & 126.51 & M6 & 11.10 & 11.27 & 0.046 & 0.055\\
        8 & 164409359522965120 & 128.03 & M5 & 11.90 & 12.08 & 0.045 & 0.054\\
        9 & 150908490604475520 & 132.86 & M5 & 12.00 & 12.18 & 0.045 & 0.053\\
        10 & 146487560507840768 & 123.63 & M4 & 10.76 & 10.93 & 0.045 & 0.052\\
        11 & 164550882989640192 & 117.39 & M2 & 9.43 & 9.58 & 0.045 & 0.052\\
        12 & 164513022853468160 & 124.69 & M6 & 11.22 & 11.39 & 0.043 & 0.051\\
        13 & 163177116226018944 & 129.18 & M5 & 10.89 & 11.07 & 0.043 & 0.050\\
        14 & 147523605402800256 & 119.46 & M2 & 9.71 & 9.86 & 0.043 & 0.050\\
        15 & 162535345034688768 & 129.35 & M2 & 10.53 & 10.68 & 0.043 & 0.049\\
        16 & 164470794735041152 & 135.11 & M6 & 12.12 & 12.30 & 0.042 & 0.049\\
        17 & 151793082068521856 & 127.01 & K8 & 9.88 & 10.01 & 0.042 & 0.048\\
        18 & 151373820245230080 & 129.60 & M4 & 9.81 & 9.98 & 0.041 & 0.048\\
        19 & 164705368668853120 & 132.21 & M2 & 10.09 & 10.24 & 0.041 & 0.048\\
        20 & 148037764527442944 & 128.20 & K5 & 9.75 & 9.91 & 0.041 & 0.047\\
        21 & 152362491654557696 & 133.03 & M4 & 10.63 & 10.80 & 0.040 & 0.047\\
        22 & 152917298349085824 & 138.85 & M7 & 11.09 & 11.25 & 0.040 & 0.047\\
        23 & 164676575208109568 & 131.81 & M4 & 10.65 & 10.82 & 0.040 & 0.046\\
        24 & 3412003903495181440 & 134.50 & M2 & 10.50 & 10.65 & 0.040 & 0.046\\
        25 & 147831571737487488 & 130.37 & K7 & 9.54 & 9.68 & 0.040 & 0.046\\
        26 & 3314299238667410176 & 145.96 & M7 & 11.39 & 11.56 & 0.040 & 0.046\\
        27 & 49366530195371392 & 137.88 & M5 & 11.11 & 11.28 & 0.040 & 0.045\\
        28 & 46008862202068480 & 130.41 & M1 & 10.14 & 10.29 & 0.039 & 0.045\\
        29 & 164422961683000320 & 125.20 & M1 & 9.21 & 9.35 & 0.039 & 0.045\\
        30 & 163246832135164544 & 127.50 & M3 & 9.74 & 9.89 & 0.039 & 0.045\\
        31 & 148141775750936960 & 142.86 & M5 & 10.95 & 11.11 & 0.038 & 0.045\\
        32 & 3314328822401984768 & 142.40 & M4 & 11.30 & 11.47 & 0.038 & 0.044\\
        33 & 3314352530621527296 & 147.83 & M5 & 11.14 & 11.32 & 0.037 & 0.044\\
        34 & 164666022471759232 & 129.82 & K6 & 9.34 & 9.49 & 0.037 & 0.044\\
        35 & 164504467278644096 & 129.30 & K8 & 9.38 & 9.53 & 0.037 & 0.044\\
        36 & 147373010964871040 & 148.08 & M6 & 11.29 & 11.48 & 0.036 & 0.043\\
        37 & 3313476524795016448 & 142.52 & M1 & 10.08 & 10.23 & 0.036 & 0.043\\
        38 & 3313414750283302400 & 150.90 & M4 & 11.05 & 11.22 & 0.035 & 0.043\\
        39 & 148420639387738112 & 147.51 & M7 & 10.74 & 10.91 & 0.034 & 0.043\\
        40 & 3408923003195121024 & 159.73 & M5 & 12.02 & 12.20 & 0.034 & 0.043\\
        41 & 147614422487144960 & 158.74 & M7 & 11.34 & 11.51 & 0.034 & 0.042\\
        42 & 151374198202645376 & 137.57 & M0 & 9.38 & 9.53 & 0.033 & 0.042\\
        43 & 156842486140929024 & 160.86 & M5 & 11.69 & 11.87 & 0.033 & 0.042\\
        44 & 145494460991086976 & 161.02 & M6 & 12.13 & 12.31 & 0.032 & 0.042\\
        45 & 151286550806099712 & 158.28 & M3 & 11.22 & 11.39 & 0.031 & 0.041\\
        46 & 145210099794710272 & 160.59 & M2 & 11.02 & 11.19 & 0.031 & 0.041\\
        47 & 164518589131083136 & 129.36 & K3 & 8.19 & 8.25 & 0.031 & 0.041\\
        48 & 3314258999116827776 & 143.66 & K7 & 9.17 & 9.30 & 0.030 & 0.041\\
        49 & 152284735566828032 & 168.02 & M6 & 12.29 & 12.46 & 0.030 & 0.041\\
        50 & 3406540769517899392 & 160.03 & M2 & 10.32 & 10.47 & 0.030 & 0.040\\
        51 & 3419134438264747648 & 170.04 & K6 & 11.79 & 11.92 & 0.030 & 0.040\\
        52 & 147727672184672640 & 161.02 & M3 & 10.82 & 10.98 & 0.030 & 0.040\\
        53 & 145213295250374016 & 160.88 & M2 & 10.76 & 10.92 & 0.030 & 0.040\\
        54 & 145225596036660224 & 162.72 & M0 & 10.25 & 10.40 & 0.030 & 0.040\\
        55 & 3414676232147787136 & 169.59 & M5 & 11.68 & 11.86 & 0.030 & 0.039\\
        56 & 145209442664192896 & 162.32 & M2 & 10.70 & 10.89 & 0.030 & 0.039\\
        57 & 145203811962545152 & 162.36 & M1 & 10.55 & 10.71 & 0.030 & 0.039\\
        58 & 145212711134828672 & 164.18 & M5 & 10.85 & 11.01 & 0.030 & 0.039\\
        59 & 3314338163954491136 & 164.36 & M1 & 9.28 & 9.42 & 0.027 & 0.039\\
        60 & 3419186939943738880 & 175.94 & M4 & 10.80 & 10.97 & 0.026 & 0.039\\
        61 & 145157941711889536 & 166.61 & K7 & 9.36 & 9.50 & 0.025 & 0.038\\
        62 & 156902512603777408 & 165.22 & K6 & 8.97 & 9.10 & 0.024 & 0.038\\

		\hline
	\end{tabular}
		\caption{Same as in table \ref{sec:BPICStars} but for the TAA sample}
	\label{sec:TAAStars}

\end{table*}

\section{Cumulative average evolution of group specific detection probabilities maps}


\bsp	
\label{lastpage}
\end{document}